\RequirePackage{fix-cm}
\documentclass[twocolumn]{svjour3}          
\smartqed  
\usepackage{graphicx}
\usepackage{times}
\usepackage{latexsym}
\usepackage{url}
\usepackage{graphicx}
\usepackage{dblfloatfix}
\usepackage{amssymb}
\usepackage{multirow}
\usepackage{amsmath}
\usepackage{tabulary}
\usepackage{booktabs}
\usepackage{makecell}
\usepackage[dvipsnames]{xcolor}
\usepackage{comment}
\usepackage{caption}
\usepackage{xcolor}
\usepackage{hyperref}
\usepackage{diagbox}

\begin{document}

\title{Evaluating BERT-based Scientific Relation Classifiers for Scholarly Knowledge Graph Construction on Digital Library Collections
\thanks{\textbf{Grants}: This material is based upon work supported by the National Science Foundation under Grant No. OAC 1939929 and by the European Research Council for the project ScienceGRAPH (Grant agreement ID: 819536). \\
 \textbf{Conflicts of interest}: Not applicable \\
 \textbf{Data/Code availability}: The clean corpora and the source codes of selected classification models are publicly available. The corresponding noisy corpora and related codes will be released after the paper is published.\\
 \textbf{Ethics approval}: Not applicable \\
 \textbf{Consent to participate}: Not applicable \\
 \textbf{Consent for publication}: Not applicable
 }
}

\titlerunning{Evaluating BERT-based Scientific Relation Classifiers}        

\author{Ming Jiang        \and
        Jennifer D’Souza  \and
        S\"{o}ren Auer    \and
        J. Stephen Downie
}

\authorrunning{Jiang et al.} 

\institute{M. Jiang \at
              University of Illinois at Urbana-Champaign, USA \\
              \email{mjiang17@illinois.edu}           
           \and
           J. D’Souza \at
              TIB Leibniz Information Centre for Science and Technology \\
              Hannover, Germany
           \and
           S. Auer \at
              TIB Leibniz Information Centre for Science and Technology \\
              L3S Research Center at Leibniz University of Hannover \\ Hannover, Germany
           \and
           J.S. Downie \at
              University of Illinois at Urbana-Champaign, USA
}

\date{Received: date / Accepted: date}

\maketitle

\begin{abstract}
The rapid growth of research publications has placed great demands on digital libraries (DL) for advanced information management technologies. To cater to these demands, techniques relying on knowledge-graph structures are being advocated. In such graph-based pipelines, inferring semantic relations between related scientific concepts is a crucial step. Recently, BERT-based pre-trained models have been popularly explored for automatic relation classification. Despite significant progress, most of them were evaluated in different scenarios, which limits their comparability. Furthermore, existing methods are primarily evaluated on clean texts, which ignores the digitization context of early scholarly publications in terms of machine scanning and optical character recognition (OCR). In such cases the texts may contain OCR noise, in turn creating uncertainty about existing classifiers' performances. To address these limitations, we started by creating OCR-noisy texts based on three clean corpora. Given these parallel corpora, we conducted a thorough empirical evaluation of eight \textsc{Bert}-based classification models by focusing on three factors: (1) \textsc{Bert} variants; (2) classification strategies; and, (3) OCR noise impacts. Experiments on clean data show that the domain-specific pre-trained \textsc{Bert} is the best variant to identify scientific relations. The strategy of predicting a single relation each time outperforms the one simultaneously identifying multiple relations in general. The optimal classifier’s performance can decline by around 10\% to 20\% in F-score on the noisy corpora. Insights discussed in this study can help DL stakeholders select techniques for building optimal knowledge-graph-based systems.

\end{abstract}

\section{Introduction}
\label{intro}
Digital libraries (DL) play an important role in promoting scholarly knowledge dissemination and exploration, which is a critical part of scholarly communication. While the accessibility of massive scholarly records in DL brings practical benefits to a variety of research communities for scholarship development, this phenomenon presents new challenges to digital librarians and data curators in managing scholarly information from various sources. In particular,  
existing modes of document-based scholarly communication create challenges for researchers looking to obtain comprehensive, fine-grained and context-sensitive scholarly knowledge for their specific research topics, especially for multi-disciplinary research~\cite{Jaradeh2019ORKG}. According to \cite{auer2018towards,auer2019orkg}, current keyword-based methods for indexing scholarly articles cause related documents to be scattered and emphasize concentration on selected keywords rather than all aspects of knowledge in an article. Given these limitations, in order to optimize scholarly knowledge organization and representation in DL, some initiatives~\cite{Jaradeh2019ORKG,stephen} advocate for building an interlinked and semantically rich knowledge graph structure using a combination of human curation and machine learning techniques.

In the process of building knowledge graphs based on scholarly publications, one of the key steps requires establishing semantic relationships between identified scientific terms. Given that paraphrasing is a common phenomenon in natural languages, many identified semantic relationship instances essentially indicate the same relation by different expressions, which can lead to information redundancy in the constructed graphs over large corpora. To avoid this issue, the task of classifying scientific relations (i.e., identifying the appropriate relation type for each related concept pair from a set of predefined relations) is indispensable. Recently, some researchers in the natural language processing (NLP) community have defined seven regular scientific relation types based on their empirical annotations on scholarly articles ~\cite{augenstein2017semeval,gabor2018semeval}, which include \textsc{Hyponym-Of}, \textsc{Part-Of}, \textsc{Usage}, \textsc{Compare}, \textsc{Conjunction}, \textsc{Feature-Of}, and \textsc{Result}. The annotations are in the form of generalized relation triples: $\langle$experiment$\rangle$ \textsc{Compare} $\langle$another experiment$\rangle$; $\langle$method$\rangle$ \textsc{Usage} $\langle$data$\rangle$; $\langle$method$\rangle$ \textsc{Usage} $\langle$re- search task$\rangle$. 

In this age of the ``deep learning tsunami'', prior work has started to take advantage of powerful neural network techniques to build scientific relation classifiers for the improvement of performance ~\cite{manning2015computational}. With the recent introduction of transformer models such as \textsc{Bert} (i.e., Bidirectional Encoder Representations from Transformers)~\cite{bert} models, the opportunity to obtain boosted machine learning systems is further accentuated. 
While prior works~\cite{scibert,mre19} have achieved high classification performance, the results reported in these studies are somewhat incomparable because the proposed methods are usually evaluated on different evaluation corpora. This issue leads to a difficulty in obtaining comparable results and conclusive insights about the effectiveness of developed classifiers in real-world practice. In particular, in the context of academic DLs, digital librarians might find it hard to select the optimal toolkit to satisfy their needs for scholarly knowledge organization based on the findings of prior evaluations (e.g., the underlying data could be different in content, scale, and diversity). Moreover, existing studies focus primarily on the development of techniques for scientific relation classification in the context of unscanned clean corpus ~\cite{scibert,luan2018multi,gabor2018semeval}, suggesting that researchers tend to concentrate on the ability of machines to automatically identify scientific concepts' relationships from recent scholarly publications that are born in the digital format. However, in broader practice, there exists a wide range of DL collections that were originally published in print and later digitized by machine scanning and OCR. These texts inevitably involve unique digitization noise such as OCR errors, which could challenge the advanced learning-based relation classification techniques. This gap raises further uncertainties about the robustness of state-of-the-art classification models to handle text noise caused by digitization when making predictions.

To address the aforementioned limitations, we focused on providing a comprehensive examination of state-of-the-art BERT-based classification models in a comprehensive and comparable environment considering both clean and OCR-noise data scenarios. With the goal of helping stakeholders within DL select the optimal tool for building scholarly knowledge graphs, we propose a two-stage examination pipeline to: (1) identify the optimal model setting for clean texts; and, (2) further investigate the resilience of the identified optimal model setting for noisy texts with OCR errors, which is one of the most universal limitations in DL collections. Regarding the model setting, we particularly focused on the impact of two key factors: (1) classification strategies (i.e., predicting either a single relation or multiple relations at one time); and, (2) \textsc{Bert} model variants with respect to the domain (i.e., generic or scientific texts) and vocabulary case (i.e., cased or uncased) of their pre-training corpus. As to the measurement of the optimal model's resilience to OCR noise, we started with generating aligned noisy-clean instance pairs by randomly adding a controlled ratio of OCR errors into each clean instance based on an existing word-level OCR-error dictionary, which was provided by HathiTrust Research Center~\cite{ocr_dict}. Given the parallel data, we then compared the optimal model's performance differences on noisy versus clean texts. 

Considering that real-world DLs may have various data settings and that such diversity would probably influence the model's performance, we assessed the performance of each model on three corpora including: (1) a single-domain corpus with sparse relation annotations on scholarly publication abstracts in the NLP area~\cite{gabor2018semeval}; (2) a multiple-domain corpus covering more abundant relations annotated on the publication abstracts from various artificial intelligence (AI) conference proceedings~\cite{luan2018multi}; and, (3) a combination of previous two corpora where the distribution of data domains are unbalanced and annotations are provided by two different groups of annotators. The motivation in building this corpus was to simulate real data settings in digital libraries. With a similar concern, we prepared two noisy versions of each clean corpus with an emphasis on the amount of OCR errors, which included a low-noise version (~18\%) and a high-noise version (~49\%). 


Experiments on three different clean evaluation corpora showed that the  uncased \textsc{Bert} model pre-trained on scholarly domain-specific texts  was the best variant to classify the type of scientific relations. Regarding the classification strategies, in general, the strategy of predicting a single relation each time achieved a higher classification accuracy than the one identifying multiple relation types simultaneously. By further examining the optimal resulting classifier's performances on three corresponding noisy corpora, we found that the classifier's predictability clearly decreased between 10\% to 20\% in F-score, indicating that the decreasing rate is even higher with the increasing of the amount of OCR errors in the corpus.


In summary, following our examination pipeline, we addressed the following research questions in this study: 

\begin{enumerate}
    \item \textbf{RQ\theenumi}: What is the performance of eight BERT-based classifiers for scientific relation classification on clean data?
    \item \textbf{RQ\theenumi}: Which of the seven studied relation types were easy to identify by the selected optimal classifiers?
    \item \textbf{RQ\theenumi}: What kinds of prediction errors are frequently made?
    \item \textbf{RQ\theenumi}: How do OCR errors impact the overall optimal classifier's performances?
    \item \textbf{RQ\theenumi}: Which of the seven studied relation types are more robust to OCR errors?
\end{enumerate}

 \section{Related Work}
 \label{sec:related}
\subsection{Relations Mined from Scientific Publications}
Overall, knowledge is organized in digital libraries based on two main aspects of a digital collection: (1) metadata; and, (2) content ~\cite{dlko,kglib}. The latter aspect can be further divided into free-form content and ontologized content (i.e., axiomatically defined formal content). In this context, the main categories of relations explored in scholarly publications included two groups. 

One group includes metadata relations such as authorship, co-authorship, and citations~\cite{meta,coauthorship}. Research in this group has focused mainly on examining the social dimension of scholarly communications, such as research impact assessments~\cite{researchimpact,impact2,re_assess_1,re_assess_2,re_assess_3,re_assess_4,re_assess_new_5,re_assess_new_6}, co-author prediction~\cite{coauthorship,coauthor_1,coauthor_2,coauthor_new_1,coauthor_new_2} and scholarly community analysis~\cite{meta}. Popular data resources that have been explored in this group include Microsoft Academic Graph~\cite{microsoft} and PubMed~\cite{pubmed}.

The second group includes content-based semantic relations, either as semantic categories empirically defined on the basis of the given free-form content ~\cite{content,constituency} or as semantic categories formally defined by a systematic conceptual analysis of the properties of concepts within a subject area ~\cite{ontology,scholarontology}. In the framework of automatic systems, free-form content-based relations have been examined in terms of: (1) relation identification (i.e., recognize related scientific term pairs)~\cite{gabor2018semeval,constituency}; and (2) relation classification (i.e., determine the relation type of each term pair, where the relation types are typically pre-defined) \cite{mre19,scibert,luan2018multi}. With respect to ontology-defined properties, prior work primarily considers the conceptual hierarchy based on formal conceptual analysis~\cite{ontology,scholarontology}.

We attempt to classify free-form content-based semantic relations. Given that digital libraries are interested in the creation of linked data~\cite{hallo2016current}, our attempted task directly facilitates the creation of scholarly knowledge graphs and offers structured data to support librarians in generating linked data.

\begin{table*}[b]
\centering
\resizebox{0.95\textwidth}{!}{
\begin{tabular}{l|p{7.5cm}|cc|cc|cc}
\toprule
\multirow{2}{*}{\textbf{Id}} & \centering \multirow{2}{*}{\textbf{Relation Type}} & \multicolumn{2}{c}{\textbf{SemEval18}} & \multicolumn{2}{|c|}{\textbf{SciERC}} & \multicolumn{2}{c}{\textbf{Combined}} \\
\cmidrule(lr){3-4} \cmidrule(lr){5-6}\cmidrule(lr){7-8}
 & & Total & \% & Total & \% & Total & \% \\ 
\midrule
1  & \textbf{\textsc{Usage}}: a scientific entity that is used for/by/on another scientific entity. E.g. \textit{MT system} is applied to \textit{Japanese} & 658 & 42.13\% & 2,437 & 52.43\% & 3,095 & 49.84\% \\ 
2  & \textbf{\textsc{Feature-Of}}: An entity is a characteristic or abstract model of another entity. E.g. \textit{computational complexity} of \textit{unification} & 392 & 25.10\% & 264 & 5.68\%  & 656 & 10.56\% \\ 
3  & \textbf{\textsc{Conjunction}}: Entities that are related in a lexical conjunction i.e., with `and' `or'. E.g. videos from \textit{Google Video} and a \textit{NatGeo documentary} & - & - & 582 & 12.52\%  & 582 & 9.37\% \\ 
4  & \textbf{\textsc{Part-Of}}: scientific entities that are in a part-whole relationship. E.g. describing the processing of \textit{utterances} in a \textit{discourse} & 304 & 19.46\% & 269 & 5.79\% & 573 & 9.23\% \\ 
5  & \textbf{\textsc{Result}}: An entity affects or yields a result. E.g. With only 12 \textit{training speakers} for SI recognition , we achieved a 7.5\% \textit{word error rate} & 92 & 5.89\% & 454 & 9.77\%  & 546 & 8.79\% \\ 
6  & \textbf{\textsc{Hyponym-Of}}: An entity whose semantic field is included within that of another entity. E.g. \textit{Image matching} is a problem in \textit{Computer Vision} & - & - & 409 & 8.80\%  & 409 & 6.59\%  \\ 
7 & \textbf{\textsc{Compare}}: An entity is compared to another entity. E.g.  \textit{conversation transcripts} have features that differ significantly from \textit{neat texts} & 116 & 7.43\% & 233 & 5.01\%  & 349 & 5.62\%   \\ \midrule
\multicolumn{2}{c|}{\textbf{Overall}} & 1,562 & 100\% & 4,648 & 100\%  & 6,210 & 100\% \\ 
\bottomrule
\end{tabular}
}
\caption{
Overview of corpus statistics (also is accessible at \url{https://www.orkg.org/orkg/comparison/R38012}). `Total' and `\%' columns show the number and percentage of instances annotated with the corresponding relation over all abstracts, respectively.}
\label{table:1}
\end{table*}

\subsection{Techniques Developed for Semantic Relation Classification}
Both rule-based~\cite{snowball} and learning-based~\cite{dependency,relrnn} methods have been developed for relation classification. Traditionally, learning-based methods typically relied on hand-crafted semantic and/or syntactic features~\cite{snowball,dependency}. Among various methods, the strategy of applying distant supervision based on a knowledge database \cite{distant,distant2,distant3} has been widely used and followed for further improvement. The major advantage of this method is its benefits in solving the challenge of the lack of hand-labeled ground truth for model training.

In recent years, deep learning techniques have been popularly studied because they can more effectively learn latent feature representations for distinguishing between relations. An attention-based bidirectional long short-term memory network (BiLSTM)~\cite{relrnn} is one of the first top-performing systems that leveraged neural attention mechanisms to capture important information per sentence for relation classification. Another advanced system~\cite{luan19} leverages a dynamic span graph framework based on BiLSTMs to simultaneously extract terms and infer their pairwise relations. Aside from these neural methods considering the word sequence order, transformer-based models such as \textsc{Bert} \cite{bert} that use self-attention mechanisms to quantify the semantic association of each word to its context have become the current state-of-the-art in relation classification. In addition to the generic \textsc{Bert} models trained on books and Wikipedia, recently, Beltagy et al.~\cite{scibert} have developed \textsc{SciBert} which are \textsc{Bert} models trained on scholarly publications.

With respect to the classification strategy, prior work regularly adopted a single-relation-at-a-time classification (SRC) that identifies the relation type for an entity pair each time~\cite{relrnn,luan19,scibert}. To improve the classification efficiency, Wang et al.~\cite{mre19} designed a \textsc{Bert}-based classifier that could recognize multiple pairwise relationships at one time, which can be regarded as a multiple-relations-at-a-time classification (MRC). As opposed to prior work that emphasizes classification improvement, we focus on providing a fine-grained analysis of existing resources for selecting the proper tool to extract and organize scientific information in digital libraries.

\subsection{Relation Classification on Noisy Data}
In general, prior work concentrating on the classification of entity relations on noisy data usually focused on noisy annotations, either at entity level such as uncorrected entity boundaries \cite{gabor2018semeval,attention1} or at relation level such as wrong relation labels \cite{attention2,reinforce1,reinforce2}. Such noise usually comes from two sources: (1) the biases of human annotations caused by differences in personal understanding of the fine-grained text semantics; and (2) the errors of machine labeling based on distant supervision relying on a generic knowledge base, in which the target content's specific contextual information is missing.

To address this noise in labeling, some methods focus on developing an attention-based neural network model based on the distant supervision learning mechanism, in which  the model can learn to combine the generic structural information from existing large-scale knowledge pools with the corpus-specific semantic information, and hence, improve the robustness of relation classifiers \cite{attention1,attention2}. Otherwise, some studies \cite{reinforce1,reinforce2} propose to take advantage of a reinforcement learning strategy: to first select high-quality labeled data, and then feed the selected instances into a relation classifier for training.

Although prior work has made remarkable contributions to improve learning models' robustness in the face of noisy labeling, the text data that these studies rely on is still clean. Given that, our consideration of noisy data is different from prior work in that we concentrate on content-based noise in texts with a specific focus on the OCR content of digitized library collections. 

\subsection{Impact of OCR Errors on Downstream NLP tasks}
With the increasing popularity of applying NLP techniques to DL textual resources for macro-level computation research \cite{lib1,auer2018towards}, concerns about the reliability of NLP techniques for processing digitized library collections have recently been on the rise ~\cite{ocr2,ocr1,ocrda}. Based on our literature review, one of the major issues that challenges NLP techniques' reliability on OCR'd texts is their potential inclusion of errors resulting from the OCR process ~\cite{ocr2,ocr1,ocrda}. 

According to ~\cite{ocrtype}, common OCR errors include character exchange, separated words, joined words, and erroneous symbols. Given the ubiquity, uneven distribution, and heterogeneous nature of OCR errors, it is difficult to fully clean such text noise even using state-of-the-art OCR correction techniques. Given that, there exists an uncertainty about the performance of standard NLP techniques applied on texts with OCR errors.

Overall, existing work concentrating on the investigation of the impact of OCR errors on NLP tasks can be divided into two groups. One is based on quantitative analysis ~\cite{ocr2,ocr1}. In this group, researchers usually measure and compare the performance differences of the same NLP tool applied on the clean versus the OCR'd version of texts. The second group of studies is based on qualitative analysis. For example, ~\cite{ocrda} conducted a series of interviews with researchers to collect their feedback on using NLP techniques to analyze digital archives. Given the scholarly users' feedback, this study analyzed and summarized the impact of OCR errors on NLP techniques. In both groups of studies, a variety of NLP tasks were investigated, including tokenization, sentence segmentation, part-of-speech tagging, named entity recognition, topic modeling, document-level information retrieval, text classification, collocation, and authorial attribution. According to the findings of prior work, there is a consistent negative influence caused by uncorrected OCR on NLP tasks, some of which could even be “irredeemably harmed by OCR errors” ~\cite{ocr1}.

In this study, we extend prior work by exploring the influence of OCR errors on relation classification from scholarly publications. Specifically, we aim to provide a systematic examination of \textsc{Bert}-based relation classifiers for their performances on both clean and OCR-noisy texts.

\section{Corpora Preparation}
The source data for this study includes two publicly available datasets~\cite{gabor2018semeval,luan2018multi}, each of which contains 500 born-digital scholarly abstracts with manual annotations of: (1) scientific terms; and, (2) semantic relation type of each related term pairs. To provide a comprehensive examination of BERT-based relation classifiers for recognizing scientific relations, from scholarly publications, with an emphasis on the real-world DL scenarios, we prepared both clean and noisy corpora. The details of each version of corpora are described below.

\subsection{Clean Corpora}
Regarding the clean corpora, in addition to directly using two selected raw datasets as two experimental corpora, we combined these two corpora into a third new corpus, which offers a more realistic evaluation setting because it provides a larger, more diverse task representation. Table~\ref{table:1} shows the statistics of our experimental corpora, each of which is detailed in the following subsections.

\subsubsection{C1: The SemEval18 Corpus.}
This corpus was created for the seventh Shared Task organized at SemEval-2018~\cite{gabor2018semeval}. The data was collected from scholarly publications in the ACL Anthology \footnote{https://aclanthology.org/}: for a total of 500 abstracts, 350 of which were partitioned as training data and the remaining 150 as testing data. Originally, annotations in this corpus contained six discrete semantic relations that were defined to capture the predominant information content. Since the relation \textsc{Topic} has far fewer annotations than other types of relations, for our evaluation, we omit this relation type and consider the following five relation types: \textsc{Usage}, \textsc{Result}, \textsc{Model}, \textsc{Part-Whole}, and \textsc{Comparison}. 

\subsubsection{C2: The SciERC Corpus}
Although the second evaluation corpus SciERC ~\cite{luan2018multi} has the same number of annotated abstracts, unlike the SemEval18 corpus, this one contains diverse underlying data domains where the abstracts were taken from 12 artificial intelligence (AI) conference/workshop proceedings in five research areas: artificial intelligence, natural language processing, speech, machine learning, and computer vision. These abstracts were annotated with the following seven relations: \textsc{Compare}, \textsc{Part-of}, \textsc{Conjunction}, \textsc{Evaluate-for}, \textsc{Feature-of}, \textsc{Used-for}, and \textsc{Hyponym-Of}. Similar to \textit{C1}, this corpus was pre-partioned by the corpus creators. They adopted  a 350/50/100 train/development/testing dataset split. Comparing \textit{C2} with \textit{C1}, we found that there are five relations, excepting \textsc{Conjunction} and \textsc{Hyponym-Of}, in \textit{C2}  that are semantically identical to the relations annotated in \textit{C1}. 

\subsubsection{C3: The Combined Corpus}
Finally, this evaluation corpus was created by merging \textit{C1} and \textit{C2}. In the merging process, we renamed some relations that are semantically identical but have different labels. First, \textsc{Used-For} in \textit{C2} and \textsc{Usage} in \textit{C1} were unified as \textsc{Usage}. Further, by observing relation annotations in \textit{C1} and \textit{C2}, we found that \textsc{Result} in \textit{C1} and \textsc{Evaluate-For} in \textit{C2} essentially express a similar meaning but the arguments of these two relations were in reverse order. For example, ``[accuracy] for [semantic classification]'' is labeled as ``accuracy'' $\to$ \textsc{Evaluate-For} $\to$ ``semantic classification'' in \textit{C2}, which can be regarded as ``semantic classification'' $\to$ \textsc{Result} $\to$ ``accuracy.'' Therefore, we renamed all instances annotated with relation \textsc{Evaluate-For} in corpus \textit{C2} into \textsc{Result} by flipping their argument order. By combining 1000 total abstracts with human annotations from two resources, our third evaluation corpus presents a comparatively more realistic evaluation scenario of large and heterogeneous data.

\begin{figure*}[t]
    \centering
    \includegraphics[scale = 0.45]{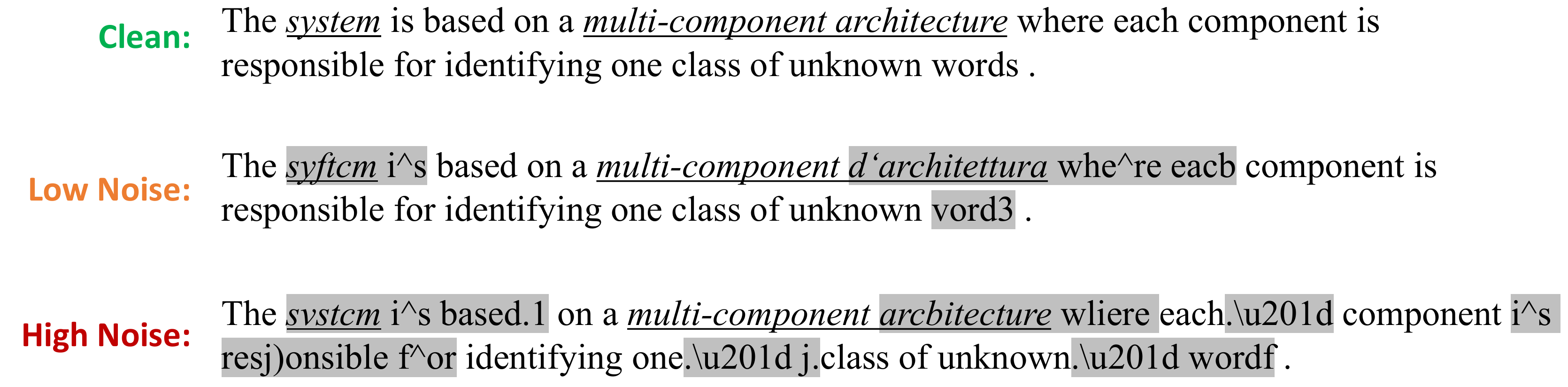}
    \caption{An illustrative example of a sentence in three versions: (1) clean; (2) low noise; and (3) high noise. \small{The underlined phrases are human-annotated scientific terms in the original non-noisy dataset; the highlighted phrase in grey is the erroneous OCR introduced text.}}
    \label{fig:noisy_exp}
\end{figure*}

\begin{table}[b]
\centering
\resizebox{0.40\textwidth}{!}{
\begin{tabular}{lcc}
\toprule
          & Low Noise & High Noise \\
\midrule
SemEval18 & 18.09\% $\pm$ 0.073  & 49.23\% $\pm$ 0.103 \\
SciERC    & 17.75\% $\pm$ 0.070  & 48.88\% $\pm$ 0.099 \\
Combined  & 17.96\% $\pm$ 0.071  & 48.97\% $\pm$ 0.100 \\
\bottomrule
\end{tabular}
}
\caption{Distribution of OCR errors in each noisy corpus}
\label{table:noisy_dist}
\end{table}

\subsection{Noisy Corpora with OCR Errors}
Given that existing work on scientific relation classification focuses primarily on developing techniques using human-cleaned plain texts, so far as we know, there appears to be no prior work that has investigated the influence of OCR errors on scientific relation classification by utilizing NLP techniques. Given that, one major challenge for us to conduct this study is the lack of available data resources. Ideally, such datasets should satisfy two demands: (1) contain both clean and noisy versions of texts that are parallel in terms of content; and (2) have human-annotated relation types on clean data as ground truth. 

Considering the high cost of directly preparing such a corpus in terms of both time consumption and human labor for tasks such as manual checking of text alignment and labeling relation types, we proposed an alternative strategy for this study that took advantage of the above off-the-shelf clean corpora with human annotations and replaced parts of their content with uncorrected OCR texts. 

To employ this strategy, we adopted an existing token-level OCR-error dictionary \cite{ocr_dict} that lists a wide range of frequently uncorrected OCR'd tokens and their corresponding corrections. According to \cite{ocr_dict}, these token pairs were collected from large-scale digitized English-language literature (178,381 volumes) published from 1700 to 1922. The total vocabulary size of this dictionary is 43,955. In order to provide a fine-grained analysis of the impact of OCR errors on \textsc{Bert}-based classifiers' performances with respect to the amount of text noise, rather than replacing all overlapping words between each scholar abstract and dictionary, we set a parameter to control the ratio of replacement. Moreover, considering the real-world uncorrected OCR in scanned texts usually has an irregular distribution, to simulate this pattern, for each scholarly abstract, we randomly selected a ratio value from a pre-set value group and our replacement was based on a random sampling of token candidates in the intersection of dictionary and the abstract content. In our empirical implementation, we prepared two levels of noisy corpus: (1) low noise corpus where the value of the replacement ratio per text is randomly chosen from the group \{0.4, 0.5, 0.7\}; and (2) high noise corpus where the ratio value per text is any of \{0.9, 1.0\}. Table~\ref{table:noisy_dist} shows the distribution of OCR errors in each corpus. On average, corpus with low text noise has around 18\% OCR errors in each text, while the corpus with high noise has around 49\% errors per abstract.

In order to have a better understanding of our prepared noisy corpora, we provide an illustrative example in Figure~\ref{fig:noisy_exp}. As expected, sentences with high text noise have more OCR errors than ones with low noise. As we can see, text noise can vary even for the same words/phrases. For example, in the low noise sentence, the word ``architecture" was replaced by ``\textbf{d'}archit\textbf{ettura}", while in the high noise sentence, this word was changed to ``arc\textbf{b}itecture". Given that many clean words in the dictionary have multiple versions of OCR errors, our random selection of one version for each word token's replacement is helpful in keeping the heterogeneous nature of OCR errors similar to a real-world scenario.




\section{\textsc{Bert}-based Scientific Relation Classifiers}
\textsc{Bert} ~\cite{bert} is a family of pre-trained language representations built on cutting-edge neural technology; it provides NLP practitioners with high-quality out-of-the-box language features that improve performance on many NLP tasks. These models return \textit{contextualized} word embeddings that can be directly employed as features for downstream tasks.
Further, with minimal task-specific extensions over the core \textsc{Bert} architecture, the embeddings can be fine-tuned to the task at hand with relatively little expense, in turn facilitating even greater boosts in task performance. 

In this study, we employ \textsc{Bert} embeddings and fine-tune them with two classification strategies: (1) single-relation-at-a-time classification (SRC); and (2) multiple-relation-at-a-time classification (MRC). In the remainder of this section, we first describe the \textsc{Bert} models that we employ and then introduce our fine-tuned SRC and MRC classifiers, respectively.

\subsection{Pre-trained \textsc{Bert} Variants}
\textsc{Bert} models as pre-trained language representations are available in several variants depending on: (1) model configuration parameters such as model size and pre-training tasks; and, (2) pre-training data settings such as language, vocabulary case and text domain. In this study, we selected the following four core variants based on the combination of two key factors of the pre-training corpus, each with two categories: (1) text domain (i.e., generic or scientific); and (2) vocabulary case (i.e., cased or uncased).

\paragraph{\textbf{\textsc{Bert}$_{\text{BASE}}$}}\footnote{https://github.com/google-research/bert} The first two models we use are in the category of pre-trained \textsc{Bert}$_{\text{BASE}}$. They were pre-trained on billions of words from the text data comprising the BooksCorpus (800M words)~\cite{zhu2015aligning} 
and English Wikipedia (2,500M words). Our two selected models are: (1) a \textbf{cased} model (where the case of the underlying words were preserved when training \textsc{Bert}$_{\text{BASE}}$); and, (2) an \textbf{uncased} model (where the underlying words were all lowercased when training \textsc{Bert}$_{\text{BASE}}$).

\paragraph{\textbf{\textsc{SciBert}}}\footnote{https://github.com/allenai/scibert} The next two models adopted in this study are in the category of pre-trained scientific \textsc{Bert} called \textsc{SciBert}. They are language models based on \textsc{Bert} but trained on a large corpus of scientific text. In particular, the pre-training corpus is a random sample of 1.14M papers from Semantic Scholar~\cite{ammar2018construction} consisting of the full text of published papers, 18\% from the computer science domain and 82\% from the broad biomedical domain. Like \textsc{Bert}$_{\text{BASE}}$, for \textsc{SciBert}, we use both its \textbf{cased} and \textbf{uncased} variants.

\subsection{Fine-tuned \textsc{Bert}-based Classifiers}
We implement the aforementioned \textsc{Bert} models within two neural system extensions that respectively adopt different classification strategies. 

\paragraph{\rm \textbf{Single-relation-at-a-time Classification (SRC)}} Classification models built for SRC generally extend the core \textsc{Bert} architecture with one additional linear classification layer that has $K \times H$ dimensions, where $K$ is the number of labels (i.e., relation types) and $H$ denotes the dimension of the word embedding space.  The label probabilities are further normalized by using a  softmax function, and the classifier assigns the label with the maximum probability to each related concept pair. 

\paragraph{\rm \textbf{Multiple-relations-at-a-time Classification (MRC)}}  
This strategy is a more recent innovation on the classification problem in which the classifier can be trained with all the relation instances in a sentence at a time or predict all the instances in one pass, as opposed to separately for each instance. In this case, however, the core \textsc{Bert} architecture's self-attention mechanism is modified to efficiently consider the representations of the relative positions of scientific terms~\cite{mre19}, which makes the encoding of the novel multiple-relations-at-a-time problem affordable. To obtain the classification probabilities, similar to the SRC strategy, the MRC is extended with a linear classification layer. However, at this time, this layer focuses on simultaneously calculating the probability per label assigned to each related term pair in a sentence. Finally, the label assignment per pair is the same as the one in SRC based on a softmax function. 

\begin{table*}[!t]
\centering
\resizebox{1.0\textwidth}{!}{
    \begin{tabular}{l c c| c c| c c || c c| c c| c c ||c c }
    \toprule
\multirow{3}{*}{} & \multicolumn{6}{c ||}{\textbf{SRC}}                                                                   & \multicolumn{6}{c ||}{\textbf{MRC}}                                                                   & \multicolumn{2}{c}{\multirow{2}{*}{\textbf{Avg$\pm$Std}}} \\ \cmidrule{2-13}
                  & \multicolumn{2}{c |}{SemEval18} & \multicolumn{2}{c |}{SciERC} & \multicolumn{2}{c ||}{Combined} & \multicolumn{2}{c |}{SemEval18} & \multicolumn{2}{c |}{SciERC} & \multicolumn{2}{c ||}{Combined} & \multicolumn{2}{l}{}                           \\
                  & Acc.           & F1           & Acc.          & F1         & Acc.           & F1          & Acc.           & F1           & Acc.          & F1         & Acc.           & F1          & Acc.                    & F1                   \\ \hline \midrule
     Bert-base uncased & 76.42 & 71.74 & 84.6 & 77.25 & 81.75 & 77.38 & 80.4 & 79.98 & 83.42 & 74.84 & 80.84 & 76.29 & 81.24$\pm$2.84 & 76.25$\pm$2.78 \\  
    Bert-base cased & 73.58 & 71.14 & 85.32 & 77.92 & 78.73 & 74.38 & 79.55 & 78.44 & 83.72 & 75.07 & 79.42 & 74.8 & 80.05$\pm$4.14 & 75.29$\pm$2.65 \\ \midrule
    Scibert cased & 73.58 & 69.72 & \textbf{86.86} & \textbf{79.65} & \textbf{84.46} & \textbf{81.60} & 80.11 & 78.32 & 83.42 & 74.35 & \textbf{81.80} & \textbf{77.68} & 81.71$\pm$4.60 & 76.89$\pm$4.25\\ 
    Scibert uncased & \textbf{80.97} & \textbf{79.42} & 86.14 & 79.49  & 83.11 & 80.27 & \textbf{81.82} & \textbf{80.54}  & \textbf{84.33} & \textbf{77.44} & 81.06 & 76.76 & \textbf{82.91$\pm$2.04} & \textbf{78.99$\pm$1.54}\\ \midrule
    Avg. Scores & \multicolumn{6}{c ||}{Acc. 84.10   F1 80.22}& \multicolumn{6}{c||}{Acc. 82.35 F1 77.52}\\
    \bottomrule
    \end{tabular}
}
\scriptsize
\caption{Scientific relation classification results over three datasets (SemEval18, SciERC, \& Combined), four \textsc{Bert} model variants (\textsc{Bert} cased \& uncased; \textsc{SciBert} cased \& uncased), and two classification strategies (SRC \& MRC). \footnotesize{$Acc.$ is accuracy and $F1$ is the macro F1-score; Top scores are in bold.}}
\label{tab:overalleval}
\end{table*}

\begin{figure}
    \centering
    \includegraphics[width=0.48\textwidth]{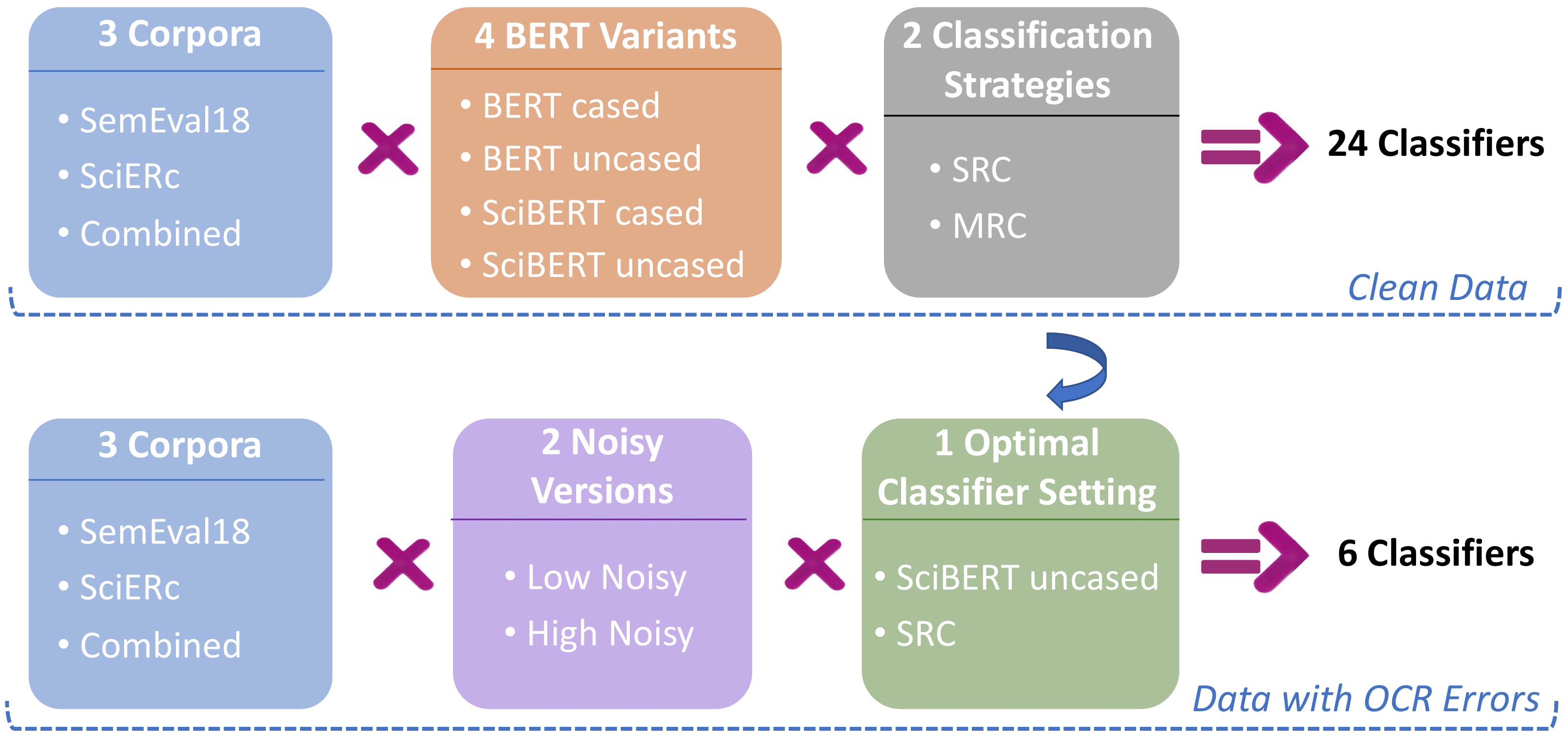}
    \caption{Classifier statistics.}
    \label{fig:classifier_sta}
\end{figure}

\section{Experiments}
\subsection{Experimental Setup}
Figure~\ref{fig:classifier_sta} provides a brief summary of classifiers investigated in this study. In total, we built \textit{24} classifiers on clean corpora and \textit{6} classifiers on the corpora with OCR errors. Each corpus had been split into training/dev/testing set by the original dataset creators. To obtain the optimal classifiers on each corpus, we tuned the learning rate parameter $\eta$ for values \{2e-5, 3e-5, 5e-5\}. For other parameters such as the number of epochs, we used default values in \textsc{SciBert} and \textsc{Bert} models.

With respect to the evaluation of classifiers' performances, we employed standard classification evaluation metrics including: Precision ($P$), Recall ($R$), F1-score ($F1$), and Accuracy ($Acc$).

\subsection{Classification Results and Analysis}
\subsubsection{RQ1: What is the performance of eight BERT-based classifiers for scientific relation classification on clean data?}

Table~\ref{tab:overalleval} provides an overview of classification results of eight classifiers trained and tested on the clean corpus based on either SRC and MRC classification strategy. Our comparison primarily focused on the following three key aspects of the classifiers.

\paragraph{\textbf{The classification strategy, i.e., SRC vs. MRC.}} Given the $Acc$ and $F1$ shown in Table~\ref{tab:overalleval}, we observed that SRC outperformed MRC on two datasets except SemEval18. One characteristic of the SemEval18 dataset is that it has a significantly lower number of annotations than the other two datasets. Given that, we infer that the novel MRC strategy is more robust than SRC because its performance level was unaffected by a drop in the number of annotations.

\paragraph{\textbf{Word embedding features, i.e., \textsc{Bert} vs. \textsc{SciBert}.}} Regarding the word embedding features encoded by different \textsc{Bert}-based models, \textsc{SciBert} outperformed \textsc{Bert} on all three corpora with higher accuracy and F1 scores. Since our experimental corpora are all scholarly data, as an expected result, word embeddings encoded by domain-specific \textsc{Bert} models can better capture the token-level semantic associations to support relation classification in the in-domain corpus than the embedding features encoded by the generic \textsc{Bert} models. 

\paragraph{\textbf{Vocabulary case in \textsc{Bert} models, i.e., cased vs. uncased.}} We observed that the uncased \textsc{Bert} models (\textsc{SciBert}: 82.91, \textsc{Bert}: 81.24) showed a higher classification accuracy than their cased counterparts (\textsc{SciBert}: 81.71, \textsc{Bert}: 80.05) on average. Further, the uncased models had an overall lower standard deviation in accuracy (\textsc{SciBert}: 2.04, \textsc{Bert}: 2.84) than the cased models (\textsc{SciBert}: 4.60, \textsc{Bert}: 4.14); comparisons on $F1$ are along similar lines. Hence, our results indicate that uncased \textsc{Bert} models can achieve more stable performances than cased variants.

In conclusion, with respect to the classification strategy, we observed that SRC outperformed MRC (see averaged scores in the last row in Table~\ref{tab:overalleval}). Nevertheless, the advanced MRC strategy demonstrates consistently robust performance that remains relatively unaffected by the size of smaller datasets compared to the SRC (e.g. SRC vs. MRC results on the SemEval18 corpus). On the other hand, with respect to \textsc{Bert} word embedding variants, from the averaged scores in the last column in Table~\ref{tab:overalleval}, the \textsc{SciBert} uncased model performed as the optimal model for encoding text features on scholarly articles. To further verify our findings, we conducted a set of statistical tests on the classification results. Considering that multiple datasets were employed in our examination and the distribution of the difference between any two samples' means may not be normally distributed, we decided to use Wilcoxon signed-rank test to explore the significance of prediction differences between any two variants of model setting. With a row-wise comparison of any two \textsc{Bert} word embedding variants’ performance (based on F1 and Acc in Table 3 respectively) over three datasets across two classification strategies, we observed that uncased SciBERT-based classifiers significantly (p$<$0.05) outperformed cased and uncased BERT-based classifiers. To explore the performance difference between SRC and MRC, we conducted a column-wise comparison over three datasets. Similarly, this comparison was based on F1 and Acc respectively. The test results showed that there was a statistically significant (p$<$0.05) difference between SRC and MRC on the testing data from SciERC and Combined corpora. Although the classification results showed that MRC provided more benefits to BERT-based classifiers than SRC in SemEval18, such performance differences lacked statistical significance.

\subsubsection{RQ2: Which of the seven studied scientific relation types were easy or challenging to be identified by the selected optimal classifiers?}
Given the optimal classifier per classification strategy for each corpus (i.e., \textsc{SciBert}-based models, either cased or uncased), we further examined these classifiers' ability to identify each type of relation labeled in the ground truth data. Tables~\ref{tab:relsemeval}, ~\ref{tab:relscierc}, and ~\ref{tab:relcomb} show the results on SemEval18, SciERC, and Combined corpus, respectively.

\paragraph{\textbf{Overview of relation type sensitivity.}} Overall, results in the three tables show that the \textsc{Usage} (\textsc{Used-For}) relation is easier to identify than other relation types under both classification strategies. One possible explanation for this observation is that \textsc{Usage} is the predominant type in all corpora, and therefore, classification models can readily learn the latent linguistic patterns of this relation type compared to the other types. 

\begin{table}[t]
\centering
\resizebox{0.48\textwidth}{!}{
    \begin{tabular}{l@{\hskip 0.8cm} c@{\hskip 0.2cm} c@{\hskip 0.2cm} c@{\hskip 0.2cm} c@{\hskip 0.2cm} c@{\hskip 0.2cm} c }
    \toprule
    \multirow{2}{*}{\thead{Relationship Type\\SemEval18}} & \multicolumn{3}{c}{\textbf{SRC}} & \multicolumn{3}{c}{\textbf{MRC}} \\ \cmidrule(lr){2-4} \cmidrule(lr){5-7} 
    & P & R & F1 & P & R & F1     \\ \hline \midrule
    \textsc{Usage} & \textbf{87.22} & 89.71 & \textbf{88.45} & 90.53 & \textbf{87.43} & \textbf{88.95} \\
    \textsc{Result} & 78.26 & \textbf{90.00} & 83.72 & \textbf{100.00} & 75.00 & 85.71 \\
    \textsc{Compare} & 85.71 & 85.71 & 85.71 & 75.00 & 85.71 & 80.00 \\
    \textsc{Model-Feature} & 66.67 & 75.76 & 70.92 & 70.83 & 77.27 & 73.91 \\
    \textsc{Part-Whole} & 79.25 & 60.00 & 68.29 & 70.83 & 72.86 & 71.83 \\
    \bottomrule
    \end{tabular}
}
\caption{Per-relation classification results of the best BERT variant under SRC and MRC strategies on SemEval18.}
\label{tab:relsemeval}
\end{table}

\begin{table}[!tb]
\centering
\resizebox{0.48\textwidth}{!}{
    \begin{tabular}{l@{\hskip 0.8cm} c@{\hskip 0.2cm} c@{\hskip 0.2cm} c@{\hskip 0.2cm} c@{\hskip 0.2cm} c@{\hskip 0.2cm} c }
    \toprule
    \multirow{2}{*}{\thead{Relationship Type\\SciERC}} & \multicolumn{3}{c}{\textbf{SRC}} & \multicolumn{3}{c}{\textbf{MRC}} \\ \cmidrule(lr){2-4} \cmidrule(lr){5-7} 
     & P & R & F1 & P & R & F1     \\ \hline \midrule
    \textsc{Used-For} & \textbf{93.30} & 91.37 & \textbf{92.32} & \textbf{88.75} & 90.24 & \textbf{89.49} \\
    \textsc{Conjunction} & 87.97 & \textbf{95.12} & 91.41 & 80.69 & \textbf{95.12} & 87.31 \\
    \textsc{Hyponym-Of} & 92.31 & 89.55 & 90.91 & 80.00 & 82.93 & 81.44\\
    \textsc{Evaluate-For} & 82.29 & 86.81 & 84.49 & 84.44 & 83.52 & 83.98 \\
    \textsc{Compare} & 72.73 & 84.21 & 78.05 & 83.87 & 68.42 & 75.36\\
    \textsc{Part-Of} & 66.04 & 55.56 & 60.34 & 65.52 & 60.32 & 62.81\\
    \textsc{Feature-Of} & 59.02 & 61.02 & 60.00 & 73.68 & 47.46 & 57.73\\
    \bottomrule
    \end{tabular}
}
\caption{Per-relation classification results of the best \textsc{Bert} variant under SRC and MRC strategies on SciERC.}
\label{tab:relscierc}
\end{table}

\begin{table}[!tb]
\centering
\resizebox{0.48\textwidth}{!}{
    \begin{tabular}{l@{\hskip 0.8cm} c@{\hskip 0.2cm} c@{\hskip 0.2cm} c@{\hskip 0.2cm} c@{\hskip 0.2cm} c@{\hskip 0.2cm} c }
    \toprule
    \multirow{2}{*}{\thead{Relationship Type\\Combined}} & \multicolumn{3}{c}{\textbf{SRC}} & \multicolumn{3}{c}{\textbf{MRC}} \\ \cmidrule(lr){2-4} \cmidrule(lr){5-7} 
    & P & R & F1 & P & R & F1     \\ \hline \midrule
    \textsc{Conjunction} & \textbf{92.56} & \textbf{91.06} & \textbf{91.80} & 85.07 & \textbf{92.68} & \textbf{88.72}\\
    \textsc{Usage}  & 91.30 & 88.98 & 90.13 & \textbf{87.96} & 87.71 & 87.84\\
    \textsc{Hyponym-Of}  & 89.39 & 88.06 & 88.72 & 83.12 & 78.05 & 80.50\\
    \textsc{Compare}  & 86.89 & 89.83 & 88.33 & 73.85 & 81.36 & 77.41\\
    \textsc{Result} & 76.36 & 75.68 & 76.02 & 84.69 & 74.77 & 79.43\\
    \textsc{Part-Of} & 75.86 & 66.17 & 70.68 & 68.33 & 61.65 & 64.82\\
    \textsc{Feature-Of}  & 58.02 & 75.20 & 65.51 & 60.28 & 68.00 & 63.91\\
    \bottomrule
    \end{tabular}
}
\scriptsize
\caption{Per-relation classification results of the best \textsc{Bert} variant under SRC and MRC strategies on the Combined corpus.}
\label{tab:relcomb}
\end{table}

For challenging relations, we found that \textsc{Feature-Of} (\textsc{Model-Feature}) and \textsc{Part-Whole} (\textsc{Part-Of}) were more difficult to identify, with lower F1 scores compared with other relation types in all three tables. Our observations could be explained by two aspects. First, there is a high level of language expression diversity in these two types of relations, which poses a difficulty for the \textsc{Bert}-based classification models to capture the linguistic patterns of these two relation types. For example, by looking into instances labeled by \textsc{Part-Whole} (\textsc{Part-Of}), we found that the key signal of this relation type included varies forms such as $\langle$A$\rangle$ ``is composed of'' $\langle$B$\rangle$, $\langle$A$\rangle$ ``...in'' $\langle$B$\rangle$, and $\langle$A$\rangle$ ``, a central instance of'' $\langle$B$\rangle$. Moreover, the comparatively lower number of annotations of these two relation types in the corpora, especially for the \textsc{SciERC} corpus, decreases the generalizability of models when predicting these two relation types, and thus leads to low F1 scores on unseen testing examples.

\paragraph{\textbf{Impact of classification strategies on per-relation type classification.}} By looking into classification strategies for the SemEval18 corpus (see Table ~\ref{tab:relsemeval}), we found that the SRC classifier and MRC classifier obtained the same classification rank order for \textsc{Usage}, \textsc{Model-Feature} and \textsc{Part-Whole}, but the opposite order for \textsc{Result} and \textsc{Compare}. In particular, the SRC classifier performed better at identifying \textsc{COMPARE}, while the MRC classifier was able to better recognize \textsc{RESULT}. As to \textsc{SciERC} (see Table ~\ref{tab:relscierc}), notably, the ability of classifying \textsc{Hyponym-Of} dropped significantly from the SRC to the MRC strategy, suggesting that the linguistic pattern of \textsc{Hyponym-Of} is hard to be captured when this relation type is mixed with other types together. 

\begin{figure*}
    \centering
    \includegraphics[scale = 0.55]{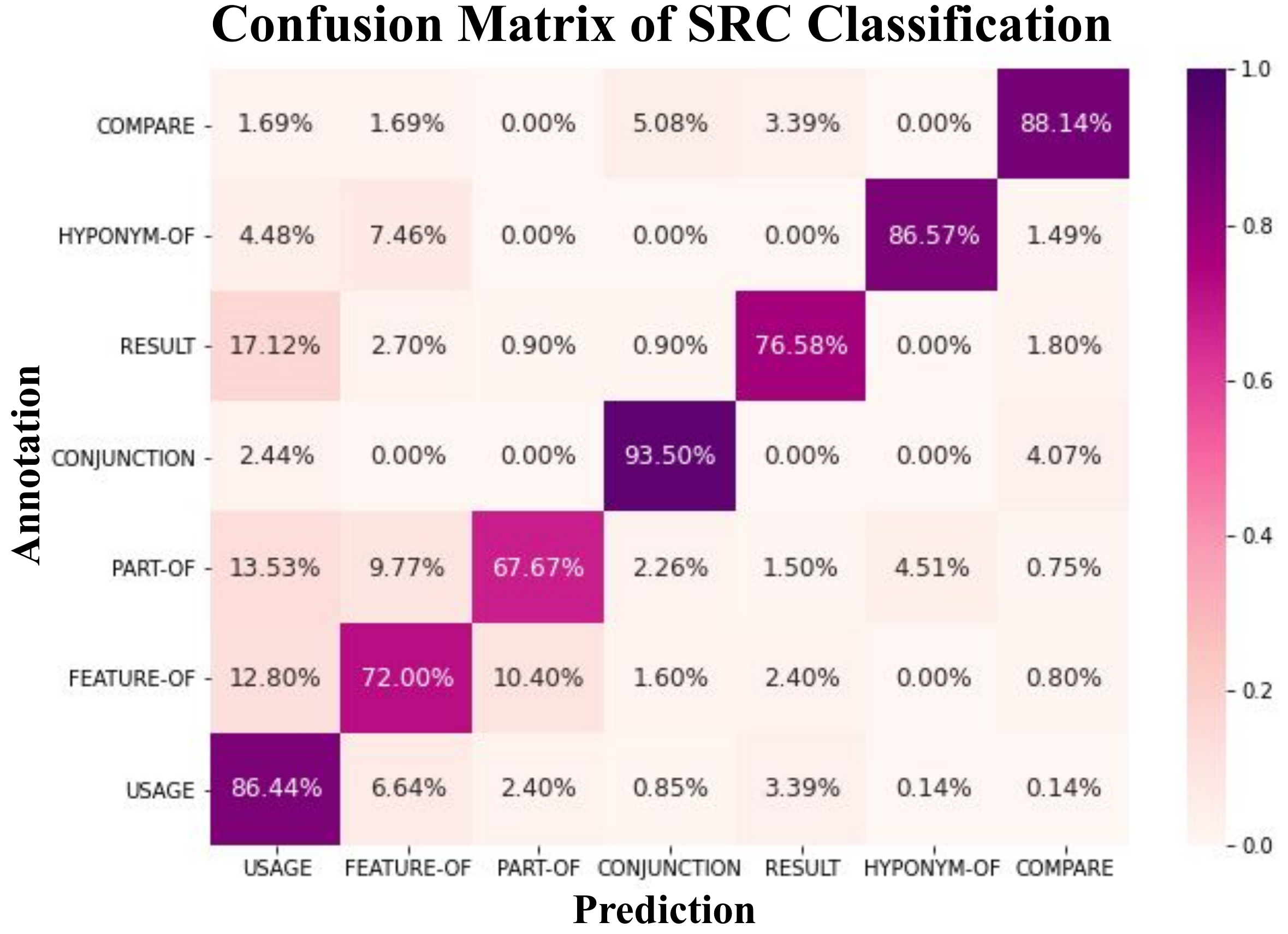}
    \caption{The confusion matrix of SRC classification on the Combined corpus.}
    \label{fig:src_cf}
\end{figure*}

\begin{figure*}
    \centering
    \includegraphics[scale = 0.55]{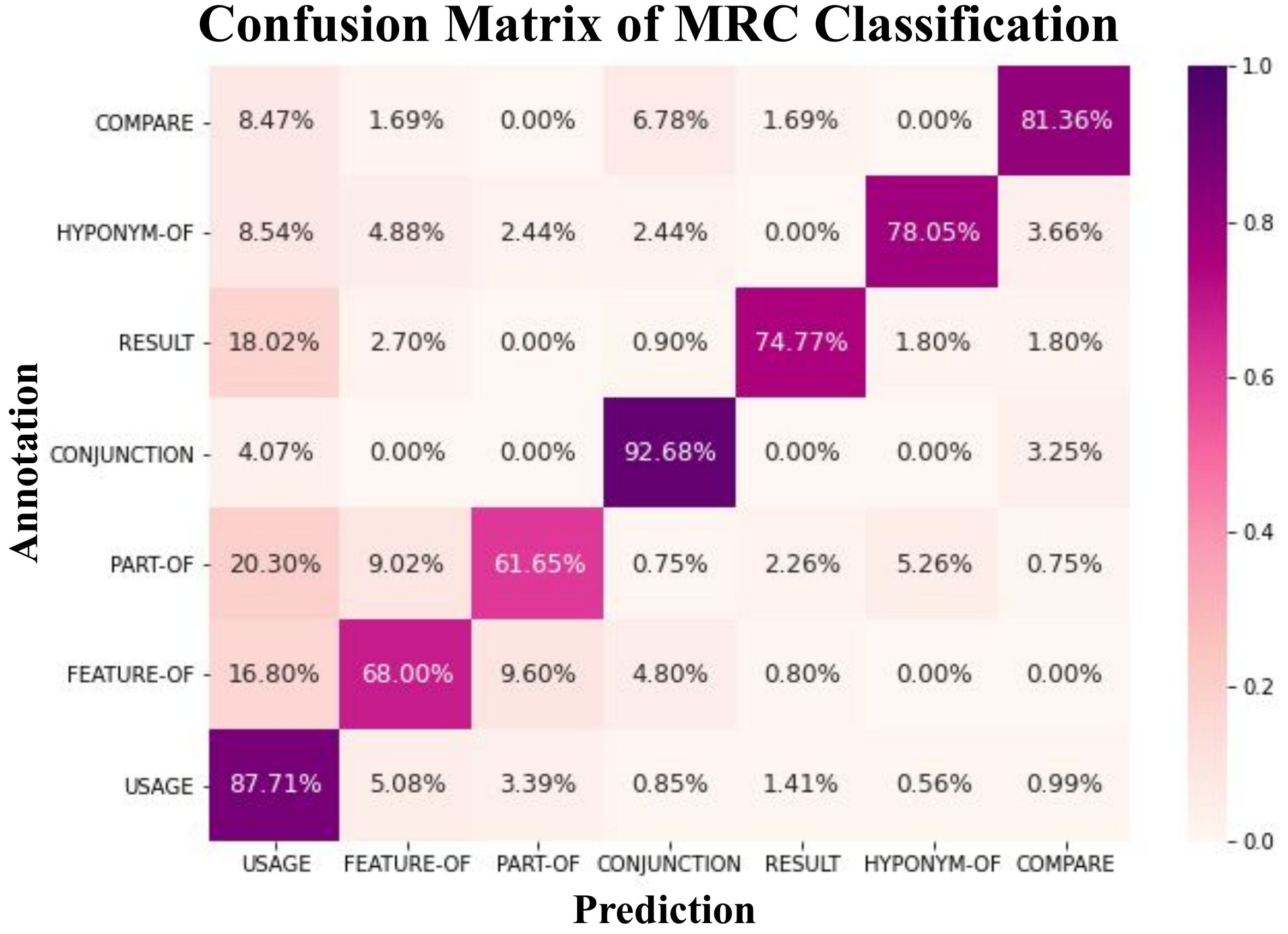}
    \caption{The confusion matrix of MRC classification on the Combined dataset.}
    \label{fig:mrc_cf}
\end{figure*}

\begin{figure*}
    \centering
    \includegraphics[width =0.75\textwidth, height=6cm]{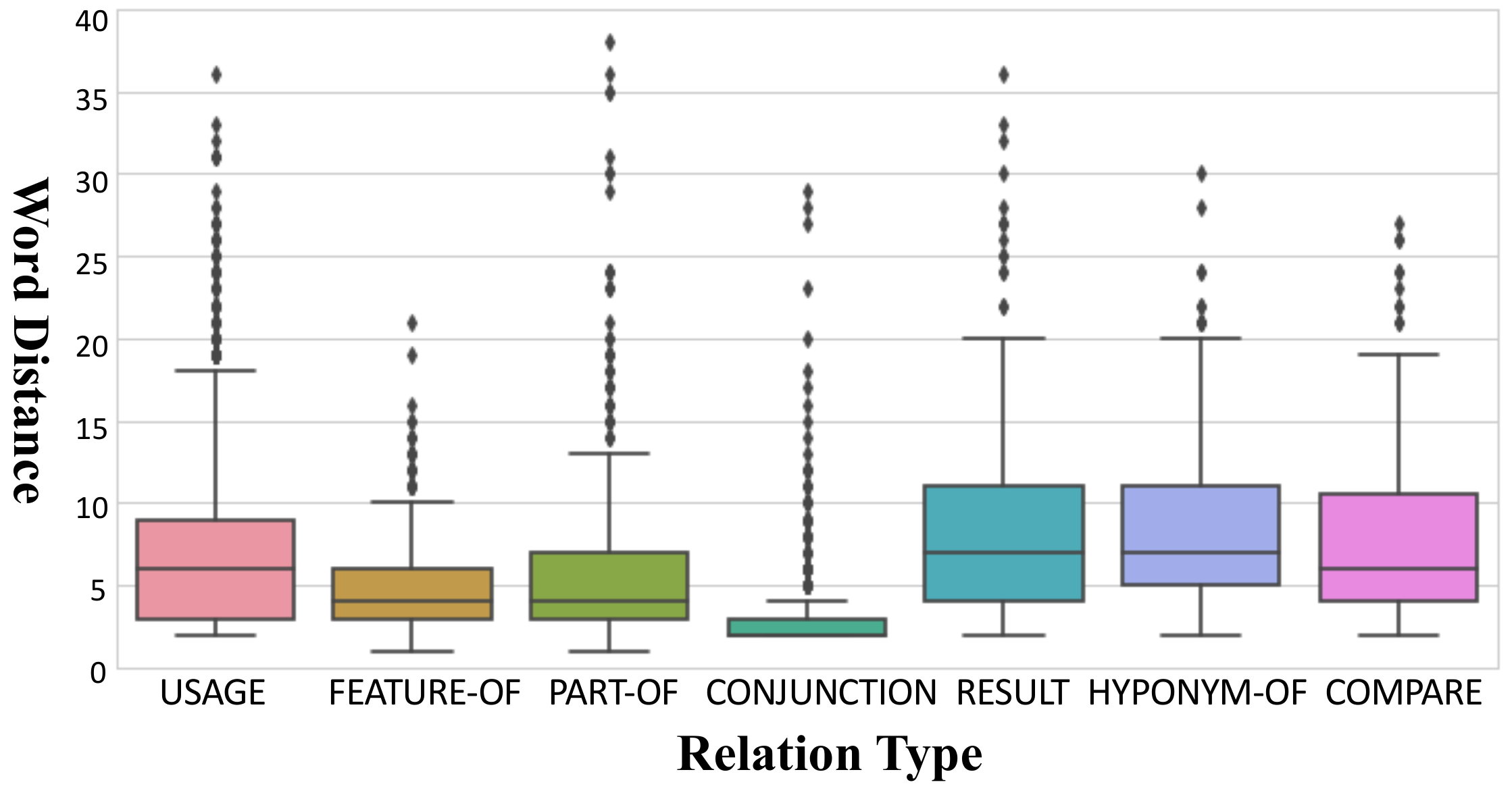}
    \caption{Distributions of word distances between scientific term pairs in abstracts in combined corpus.}
    \label{fig:worddist}
\end{figure*}

\paragraph{\textbf{Impact of heterogeneous human annotations on per-relation type classification.}} The combination of SemEval18 and \textsc{SciERC} as the Combined corpus makes this corpus more heterogeneous and realistic. Given that, we further explored whether the mix of the two groups of human annotations from different corpus sources showed any influence on the classifier's ability to identify each relation type.

Interestingly, differing from the results shown in Table~\ref{tab:relsemeval} (SemEval18) and Table~\ref{tab:relscierc} (\textsc{SciERC}) where the F1 score of predicting \textsc{Usage} (\textsc{Used-For}) ranked first in both SRC and MRC strategies, we observed that \textsc{Conjunction} was the easiest relation type to classify in the Combined corpus for both SRC and MRC classifiers (see Table~\ref{tab:relcomb}). Moreover, the F1 score of \textsc{Conjunction} in Table~\ref{tab:relscierc} is higher than the one in Table~\ref{tab:relcomb}. Since there is no \textsc{Conjunction} relation in the SemEval18 corpus, we can infer that testing examples of this relation type should be the same as the ones in the \textsc{SciERC} corpus. Given that, our results suggest that there may exist inconsistencies in the two groups of human annotations of the relation \textsc{Usage}, which influenced the model's learnability of this relation type. On the other hand, the uniform annotation of \textsc{Conjunction} from the SciERC corpus could benefit the model's ability to capture the consistent latent linguistic pattern for this relation type in the Combined corpus, and hence, allowed it to be classified more effectively. Nevertheless, we can see that the classification F1 for \textsc{Usage} was nearly the average of the scores from each individual corpus respectively. This offers a potential indicator of expected classification behavior on a heterogeneous corpus from more than one source.

\subsubsection{RQ3: What kinds of prediction errors are frequently made?}

\paragraph{\textbf{Confusion matrix of SRC and MRC classification.}} A closer look at misclassifications in the Combined corpus is depicted in the confusion matrices in Figure~\ref{fig:src_cf} for the SRC classifier and Figure~\ref{fig:mrc_cf} for the MRC classifier. Results in both figures show that classifiers tend to predominantly misclassify other relation types such as \textsc{Usage} (e.g., \textsc{Result}, \textsc{Part-Of}, and \textsc{Feature-Of}), especially in the MRC classification. One possible reason is that the \textsc{Usage} relation has the largest number of annotated instances, which caused the classifiers to be biased to this relation type. In general, unbalanced distribution of training samples (see the details in Section 3) is one of the main causes in confusion learned in machine learning systems. At the same time, the strategy of making predictions on multiple relation types simultaneously can boost such biases to learning models.

In particular, \textsc{Feature-Of}, \textsc{Part-Of}, and \textsc{Usage} are highly likely to be confused with each other in both classification strategies, suggesting that these three relationships have closer semantic associations than other relation types, and therefore making it difficult for the machine to differentiate between these three relation types. We further verified this finding by referring to the annotation guidelines of SemEval18 and \textsc{SciERC}. For example, in \textsc{SciERC}'s guideline, the instance $\langle$B ``belongs to" A$\rangle$ is semantically similar with the instance $\langle$B `` is a part of" A$\rangle$. However, the former relationship was defined as \textsc{Feature-Of}, while the latter one was a case of \textsc{Part-Of}. We also found that \textsc{Hyponym-Of} (around 6\%) is more likely to be misclassified as \textsc{Feature-Of} (in addition to \textsc{Usage}), but not vice versa. This one-sided relationship loosely demonstrates that the machine might learn a relation hierarchy between these two relation types, i.e. \textsc{Hyponym-Of} subsumes \textsc{Feature-Of}, but not the other way around. 

In conclusion, our findings show that errors in the classification of scientific relations by the optimal \textsc{Bert}-based classifier can be caused by three factors: (1) unbalanced data distribution; (2) semantic ambiguity of pre-defined relation types; and, (3) hierarchical semantics of pre-defined relation types.

\paragraph{\textbf{Word distance distribution.}} To offer another pertinent angle on the classifier error analysis, we computes the distribution of word distances between related scientific term pairs for each type of relation on the Combined corpus. The result is depicted in Figure~\ref{fig:worddist}. In general, the majority of box plots shown in Figure~\ref{fig:worddist} are skewed with a long upper whisker and a short lower whisker. This pattern indicates that the distance between paired scientific terms is typically closed in the text. As opposed to other relations, the word distance of \textsc{Conjunction} is much shorter, which makes sense because term pairs with this relationship are typically connected by a single connection term such as ``and'' and ``or''. This consistent pattern could be another reason why \textsc{Conjunction} is comparatively easier to be classified than other relations. Further, the average word distance of \textsc{Feature-Of}, \textsc{Part-Of}, \textsc{Hyponym-Of}, and \textsc{Compare} is closer to the lower quartile than the other relations. Such varied distribution may bring challenges for a classifier to identify these relations. Notably, the similar median value and spread range between \textsc{Feature-Of} and \textsc{Part-Of} could account for why they are challenging for the classification models to identify.

\subsubsection{RQ4: How do OCR errors impact the overall optimal classifier's performances?}
Following the analysis of classification results on clean corpora presented in the earlier sections, we found that the uncased \textsc{SciBert} built under SRC was optimal to identify scientific relations in general. In order to provide comparable results across corpora, in this section, we applied this optimal model and further analyzed its classification performance on the noisy version of all evaluation corpora. Our goal was to explore if there is any consistent pattern of the impact of OCR noise on predictions.


Tables~\ref{tab:ocr_semeval}, \ref{tab:ocr_sciie}, and \ref{tab:ocr_combined} show the classification results on SemEval18, SciERC, and Combined corpus, respectively. Our analysis primarily focused on two aspects: (1) the impact of OCR errors from various combinations of noisy training and testing splits in classifications; and, (2) other potential data factors, such as corpus size, associated with the impact of OCR errors. The details of each aspect is described below.

\paragraph{\textbf{Impact of OCR noise in testing data.}} Results in three tables showed that the growth of OCR errors in testing data led to a rapid drop off in the classifier's predictability performance regardless of the amount of text noise in the training data. For example, the classifier trained on the clean texts per corpus (see Table~\ref{tab:ocr_semeval}/~\ref{tab:ocr_sciie}/~\ref{tab:ocr_combined}) had a loss of F1 scores at around 5\% for the testing data with a low amount of OCR errors, and around 20\% for the one with high a high amount of text noise. Given the same training data, our observation shows that the loss of predictions increases in accordance with the growth of OCR errors in texts.

\paragraph{\textbf{Impact of OCR noise in training data.}} Regarding the number of OCR errors in the training data, in each corpus we observed that the increasing of text noise in training data helped the classifier in improving its robustness, with a lower loss of F1 score, when making predictions on testing examples, especially for the training set with a high amount of text noise. As shown in the three tables, classifiers trained on the clean data had a performance loss of around 20\% F1, while for the models built upon high noise training data, the loss decreased to 6\% F1 score on average. Our observations indicate that OCR errors in texts have an obvious impact on the performance of \textsc{BERT}-based classifier for identifying scientific relations within sentences. Interestingly, text noise in the training data has a regularization effect on the transformer-based neural network model during its learning process, which essentially benefits the model to improve its generalization ability to process the noisy unseen data.

\begin{table}[]
\resizebox{0.48\textwidth}{!}{
\begin{tabular}{@{}l||crcrcr}
\toprule
\multirow{2}{*}{\diagbox[width=7em,trim=l]{Testing}{Training}} & \multicolumn{2}{c}{Clean}                 & \multicolumn{2}{c}{Low Noise} & \multicolumn{2}{c}{High Noise} \\
\cmidrule(lr){2-3} \cmidrule(lr){4-5} \cmidrule(lr){6-7} 
                  & F1  & Loss & F1 & Loss & F1 & Loss \\
\hline \midrule
Clean             & \textbf{79.42} & -  & \textbf{75.11} & - & \textbf{64.82} & - \\
Low Noise          & 73.16  & 5.26$\downarrow$  & 72.21  & 2.90$\downarrow$  & 62.72  & 2.10$\downarrow$  \\
High Noise         & 56.35  & 23.07$\downarrow$  & 58.35  & 16.76$\downarrow$  & 58.26  & 6.56$\downarrow$  \\ 
\bottomrule
\end{tabular}
}
\caption{Classification results of uncased \textsc{SciBert} on noisy SemEval18 corpus in SRC. Top F1 scores are in bold.}
\label{tab:ocr_semeval}
\end{table}

\begin{table}[]
\resizebox{0.48\textwidth}{!}{
\begin{tabular}{@{}l||crcrcr}
\toprule
\multirow{2}{*}{\diagbox[width=7em,trim=l]{Testing}{Training}} & \multicolumn{2}{c}{Clean}                 & \multicolumn{2}{c}{Low Noise} & \multicolumn{2}{c}{High Noise} \\
\cmidrule(lr){2-3} \cmidrule(lr){4-5} \cmidrule(lr){6-7} 
                  & F1  & Loss & F1 & Loss & F1 & Loss \\
\hline \midrule
Clean             & \textbf{79.49} & -  & 77.51 & - & \textbf{78.05} & - \\
Low Noise          & 75.16  & 4.33$\downarrow$  & \textbf{77.92}  & -  & 76.35  & 1.70$\downarrow$  \\
High Noise         & 60.11  & 19.38$\downarrow$  & 69.94  & 7.57$\downarrow$  & 72.11  & 5.94$\downarrow$  \\ 
\bottomrule
\end{tabular}
}
\caption{Classification results of uncased \textsc{SciBert} on noisy SciERC corpus in SRC. Top F1 scores are in bold.}
\label{tab:ocr_sciie}
\end{table}

\begin{table}[]
\resizebox{0.48\textwidth}{!}{
\begin{tabular}{@{}l||crcrcr}
\toprule
\multirow{2}{*}{\diagbox[width=7em,trim=l]{Testing}{Training}} & \multicolumn{2}{c}{Clean}                 & \multicolumn{2}{c}{Low Noise} & \multicolumn{2}{c}{High Noise} \\
\cmidrule(lr){2-3} \cmidrule(lr){4-5} \cmidrule(lr){6-7} 
                  & F1  & Loss & F1 & Loss & F1 & Loss \\
\hline \midrule
Clean             & \textbf{80.27} & -  & \textbf{78.64} & - & \textbf{77.10} & - \\
Low Noise          & 72.84  & 7.43$\downarrow$  & 77.41  & 1.23$\downarrow$  & 74.94  & 2.16$\downarrow$  \\
High Noise         & 57.43  & 22.84$\downarrow$  & 69.29  & 9.35$\downarrow$  & 70.93  & 6.17$\downarrow$  \\ 
\bottomrule
\end{tabular}
}
\caption{Classification results of uncased \textsc{SciBert} on noisy Combined corpus in SRC. Top F1 scores are in bold.}
\label{tab:ocr_combined}
\end{table}

\paragraph{\textbf{Impact of the size of OCR noise among corpora.}} By further comparing the classification results among three corpora, we found that the predictability of classifiers on the SemEval18 corpus was much worse, with lower F1 scores than the classification models' performances on the other two corpora when the training data is noisy. For example, the predication difference could be around 10\% for each type of testing set when the training data is high noise. Given that the SemEval18 corpus has a smaller corpus size than the other two corpora, our observation suggests that large corpus size could alleviate the impact of text noise on BERT-based classification models for identifying scientific relations.

\begin{figure}[]
    \centering
    \includegraphics[width=0.48\textwidth,height=4cm]{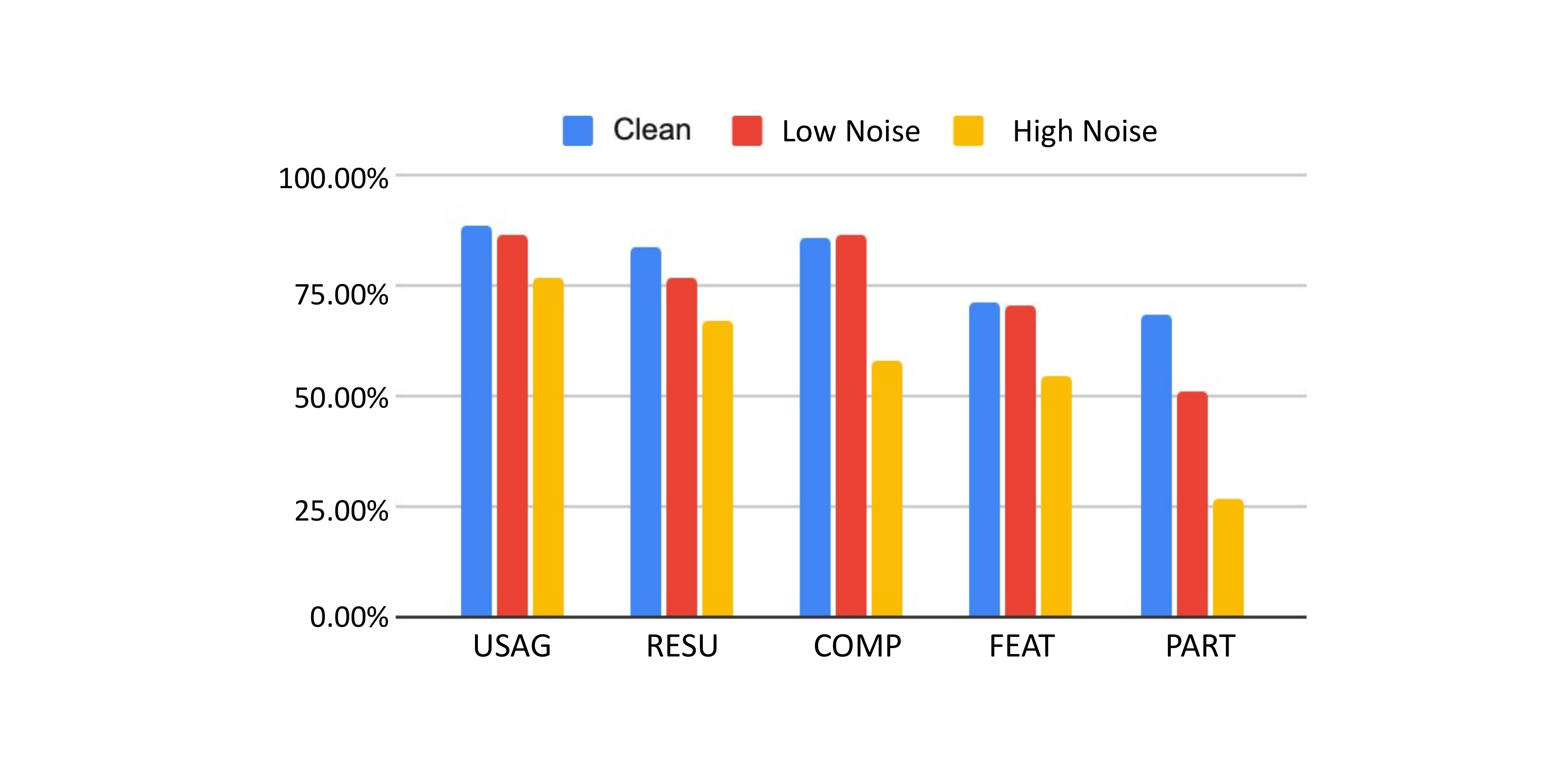}
    \caption{Per-relation type classification results on clean, low-noise and high-noise SemEval18 corpus.}
    \label{fig:ocr_semeval}
\end{figure}

\begin{figure}[]
    \centering
    \includegraphics[width=0.48\textwidth,height=4cm]{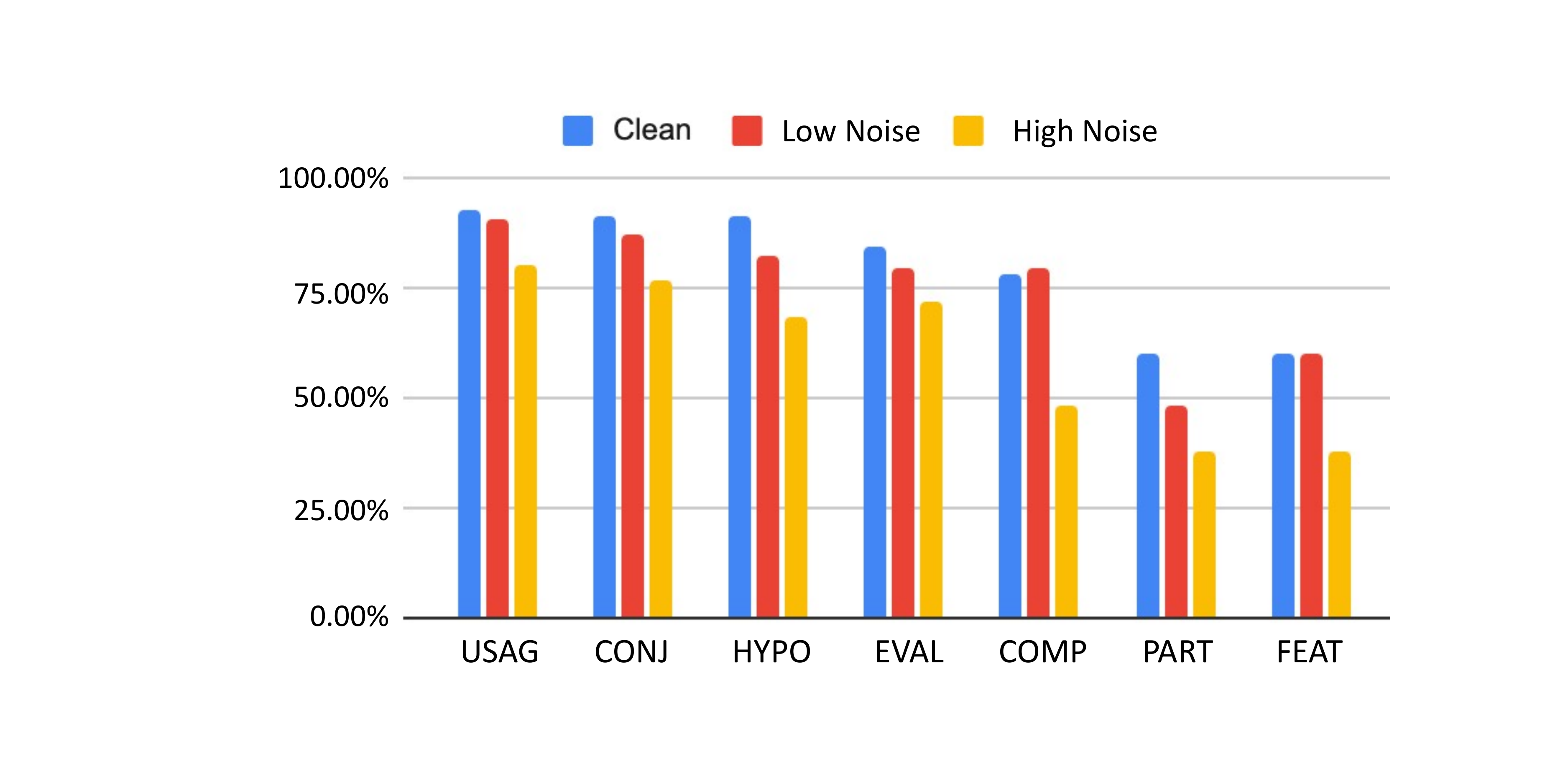}
    \caption{Per-relation type classification results on clean, low-noise and high-noise \textsc{SciERC} corpus.}
    \label{fig:ocr_scierc}
\end{figure}

\begin{figure}[]
    \centering
    \includegraphics[width=0.48\textwidth,height=4cm]{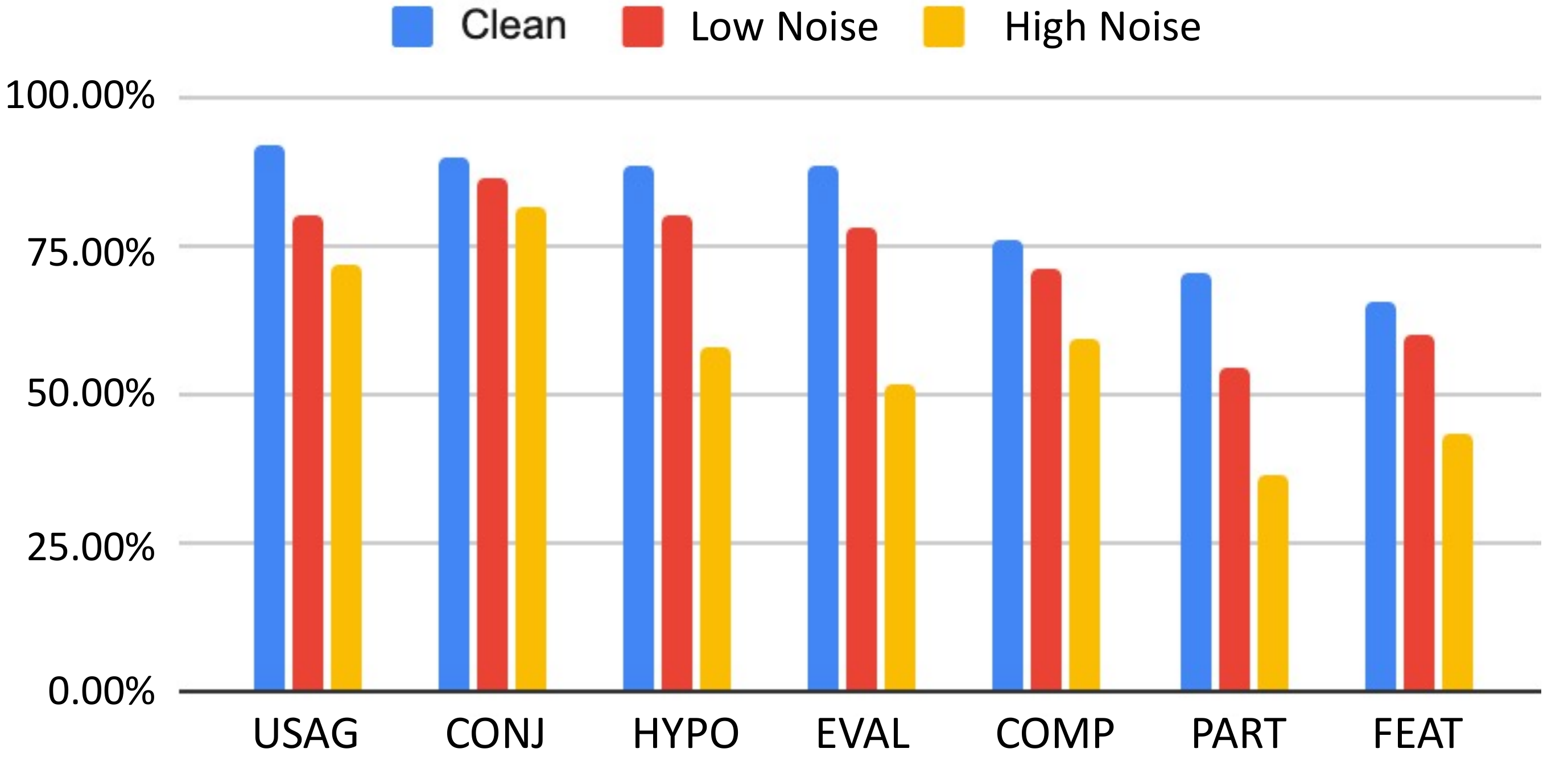}
    \caption{Per-relation type classification results on clean, low-noise and high-noise on Combined corpus.}
    \label{fig:ocr_comb}
\end{figure}

\subsubsection{RQ5: Which of the seven studied relation types are more robust to OCR errors?}
Given that the majority of state-of-the-art scientific classification techniques are developed using clean corpus, to investigate the vulnerability of such techniques to text noise with respect to each type of relations, we examined the ability of our optimal classifiers trained on the clean corpus to identify each relation type from texts with or without OCR errors. Figures~\ref{fig:ocr_semeval},~\ref{fig:ocr_scierc}, and~\ref{fig:ocr_comb} show the results on three corpora respectively.

In general, results for three corpora showed that the difficulty of the classifier to identify each relation type tends to be consistent regardless of the amount of noise in the testing texts. Comparatively, \textsc{Compare}, \textsc{Part-Of} (\textsc{Part-Whole}), and \textsc{Hyponym-Of} are three top relation types for which the classifier's performances were more sensitive to the large amount of OCR errors in terms of a high loss of F1 scores compared with other relation types, which suggests that these three types of relationships between scientific concepts are easily broken by text noise. In contrast, the predictability of each classifier on \textsc{Usage} and \textsc{Result} (\textsc{Evaluate-For}) is more robust against text noise than on other relation types. Given that, we infer that there might exist a strong semantic association between \textsc{Usage} and \textsc{Result} relations, which could overcome the disturbance of OCR errors to some extent.

By looking further into the influence of text noise by OCR error amount, we found that the difference in the classifier's ability to identify most relation types between the clean and low-noise texts is much smaller than the difference between low versus high noise texts. The only exception lies in the \textsc{Part-Of} (\textsc{Part-Whole}) relation. This result indicates that our optimal classifier can be robust to a small ratio of text noise (i.e., around 10\%) when making predictions on the majority of pre-defined relations except \textsc{Part-Of} (\textsc{Part-Whole}).

\begin{figure*}[t]
    \centering
    \includegraphics[width =1.0\textwidth]{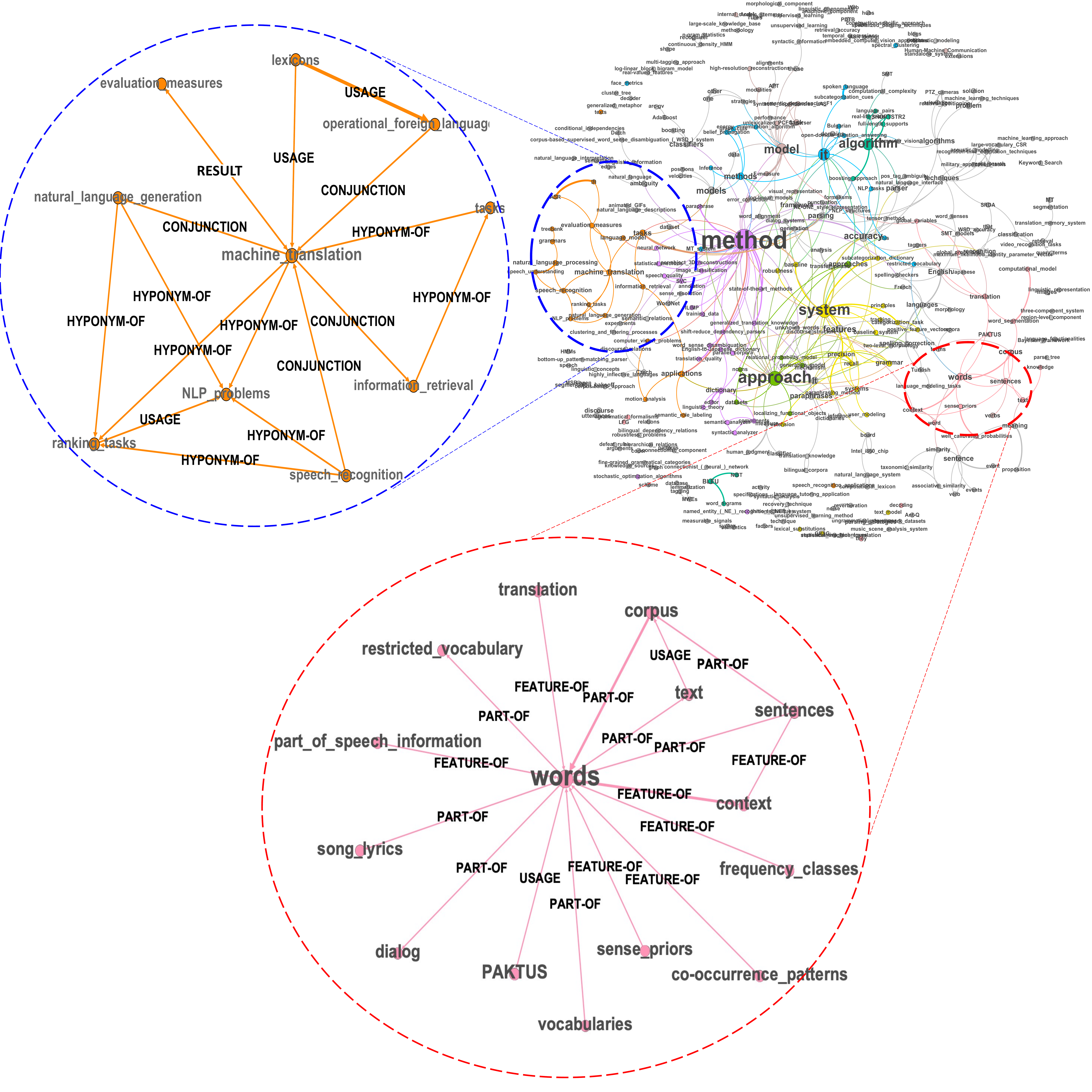}
    \scriptsize
    \caption{A knowledge graph constructed from the relation triples in the Combined corpus. Ego-network for the term ``machine\_translation'' and ``words" are presented as two illustrative examples. The node size is determined by node weighted degree. Colors denote the modularity classes based on the graph structure. \scriptsize{The graph was generated using Gephi (https://gephi.org/)}}
    \label{fig:kg}
\end{figure*}

\subsection{Use Case of Scientific Relation Classification for Scholarly Knowledge Graph Construction}
As a practical illustration of the relation triples studied in this work, finally, we built a knowledge graph from the human annotations in the Combined corpus. Figure~\ref{fig:kg} provides an illustrative example of the visualization of the resulting knowledge graph, which includes: (1) a graph at corpus-level (shown in the upper right); and, (2) two graphs at concept-level, e.g. ego-network for the concept node ``machine translation'' (shown in the upper left) and ``words'' (shown at the bottom).

Looking at the corpus-level graph, we observed that generic scientific terms such as ``method,'' ``approach,'' and ``system'' were the most densely connected nodes, as expected since these generic terms are usually found across research areas. In the zoomed-in ego-network of ``machine translation,'' we can see that \textsc{Hyponym-Of} is meaningfully highlighted by its role linking ``machine translation'' and its sibling nodes (i.e. other research tasks) including ``speech recognition,'' and ``natural language generation'' to the parent node ``NLP problems.'' The concept ``lexicon'' is usually used in (i.e., \textsc{Usage}) research on ``machine translation'' and ``operational foreign language.'' The \textsc{Conjunction} link connects the term ``machine translation'' and ``speech recognition,'' both of which aim at translating information from one source to the other one. Regarding the ego-network of ``words", we found that this term mainly involve two types of semantic relations: (1) being \textsc{Part-Of} other natural language components such as ``corpus", ``song\_lyrics", and ``sentences"; or, (2) working as features (i.e., \textsc{Feature-Of}) in NLP computations such as obtaining ``part\_of\_speech\_information" and capturing word-level ``co-occurrence\_patterns".

In addition to looking singly into the semantic association across the node as well as into link representations in the ego-network, and by further considering the network structure, we can infer the semantic hierarchy of some associated scientific concepts. For example, both ``machine translation'' are connected with ``ranking tasks'' and ``NLP problems'' by the relation \textsc{Hyponym-Of}, and ``ranking tasks'' is used for solving (i.e., \textsc{Usage}) ``NLP problems.'' This triangle structure suggests that ``ranking tasks'' is a more fine-grained hypernym to ``machine translation'' than the term ``NLP problems.'' Additionally, by comparing the structure of the ego-network of ``machine-translation" with the one of ``words", we observed the ``machine-translation"-centered graph is more dense than the ``words"-centered graph. In essence, this observation indicates that the term ``words" is a more generic concept than ``machine\_translation" because it tends to be associated with diverse independent concepts.

In summary, given the pair of related scientific concepts with the identified relation type, we can construct a knowledge graph to represent the semantic association among scientific concepts either at macro-level in terms of the whole corpus or at micro-level with regards to the fine-grained connections related to a specific concept.

\section{Discussion}
\subsection{\textsc{Bert}-based Model Recommendations for Scientific Relation Classification}
Based on the findings in this study, we provide the following recommendations to stakeholders of digital libraries for applying the optimal technique to automatically classify scientific relations from scholarly articles:

\begin{itemize}
    \item Compared with the generic domain, using corpus that specifically consists of scholarly publications to pre-train  \textsc{Bert} models benefits building classifiers in recognizing the relationship between scientific concepts.
    \item With respect to the classification strategy, SRC outperforms MRC in general. 
    \item Overall, uncased \textsc{Bert} models outperform cased ones in terms of higher accuracies and more stable performances in predictions.  
    \item Given noisy texts containing OCR errors, if the amount of text noise is small, such as around 10\%-20\%, the optimal classification model (i.e. built with uncased \textsc{SciBert} in SRC classification) trained on the clean corpus could be robust against this text noise when making predictions.
    \item For noisy texts with an uncertain amount of OCR errors, employing noisy data to train the classifiers would be helpful in building the models' generalization ability to process the texts regardless of the amount of OCR errors.
    \item In \textsc{SciBert}-based SRC classification, the large corpus size with abundant relation annotations is helpful in: (1) improving the prediction accuracy; and (2) alleviating the influence of text noise in the training data for building a classifier.
    \item Regarding relation type annotations, the large number of annotations for a relation type in the training set can help the classification model to improve its learnability on this relation type.
    \item For each pre-defined relation, the fixed syntactic structure in expressions benefits the classifier in discriminating between different relation types.
    \item Text noise is less influential on the classifier when predicting concrete scientific relations such as \textsc{Usage} and \textsc{Result} (\textsc{Evaluate-For}) compared with other relations, while this situation is contrary for some relations indicating concept hierarchies such as \textsc{Compare} and \textsc{Hyponym-Of}.
\end{itemize}

\subsection{Study Limitations and Challenges}
With a systematic review of our evaluation process, several research challenges and resulting study limitations arise, which can be summarized in three aspects: (1) human annotation collection; (2) noisy corpus preparation; and (3) evaluation methods. Each aspect is described as follows.

\subsubsection{Human Annotation Collection}
\paragraph{\textbf{Cost and resource of annotations}} As mentioned in prior work \cite{gabor2018semeval,luan2018multi}, one common challenge in research on scholarly information extraction (including both entity and relation extraction from scholarly records) is the high cost and limited resources for collecting human annotations as ground truth for training and/or testing learning-based methods. This is mainly because annotations on scientific information not only cost time and human labor similar to that involved in information extraction tasks in generic domains (e.g., newswire), they also require a higher level of annotators' expertise in the specific scientific domain of the target scholarly articles, for which it is comparatively more difficult to find adequate annotators. In addition, the expense of such annotations should be high. Due to these challenges, the number of human annotations on scientific information is usually limited, and the number of research areas involved by existing accessible human annotations is barely sufficient. 

Influenced by the aforementioned issues, our evaluation mainly focused on the corpora covering AI-related research areas, which potentially leads to some uncertainties in the performance of \textsc{Bert}-based classifiers for identifying scientific relations in other research areas such as biology and physics. 

\paragraph{\textbf{Quality of annotations as ground truth}} Following our findings, the ambiguity of relation type definitions is one of the key factors leading to the classifiers' misclassifications. There are two main aspects contributing to such ambiguity. First, the meaning of some relation types' definitions makes it difficult for humans to have a clear understanding of these relation types when annotating the data. For example, in the annotation guidelines of \textsc{SciERC} dataset ~\cite{luan2018multi}, $\langle$B$\rangle$ ``belongs to" $\langle$A$\rangle$ was defined as an instance of \textsc{Feature-Of}, while $\langle$B$\rangle$ ``is a part of " $\langle$A$\rangle$ was defined as a case of \textsc{Part-Of}. In essence, both instances are quite similar in terms of the semantic meaning. Second, different annotation systems may lead to inconsistent human annotations. For example, for two relations having a hierarchical semantic association, i.e.,  \textsc{Hyponym-Of} and \textsc{Part-Of}, the guideline of \textsc{SemEval18} dataset only considers \textsc{Part-Of}, while \textsc{SciERC}'s guidelines ask the annotators to label both relation types. Given this guideline difference, annotators are likely to annotate the real \textsc{Hyponym-Of} relationship in the \textsc{SemEval18} dataset by the tag \textsc{Part-Of}.

In addition to annotation ambiguity, there also exists annotation biases in the corpora, such as the unbalanced distribution of relation labels, which can lead to the preference of classifiers to recognize some well-represented relations(e.g., \textsc{Usage}). Due to these challenges leading to the uncertainty of annotation quality, there may exist a potential risk of underestimating the ability of \textsc{Bert}-based classifiers for identifying scientific relations in our evaluations.

\subsubsection{Digitization-based Noisy Corpus Preparation}
In order to investigate the impact of text noise caused by digitization on relation classification techniques, building an ideal noisy corpus typically requires two elements: (1) the corpus should have both clean and real-world OCR'd texts that are parallel in terms of content; and, (2) the clean corpus should have a high quality. However, it is not always possible to satisfy these two demands. One major reason is that the original version of digitized scholarly records were usually published in printed pages, which means such resources lack an off-the-shelf electronic version of texts. Given that, getting access to the fully clean texts of these scholarly resources is difficult. Besides, the aforementioned challenges of human annotation also exist in this data preparation process.

Although our presented strategy for constructing a noisy corpus in this paper can be an alternative method to address the above challenges, adding common word-level OCR noise into the clean texts based on a dictionary may not fully reflect the data patterns of the real-word digitized library collections. This limitation might involve the artifacts of engineering in our noisy corpus for further investigations.

\subsubsection{Evaluation Methods}
Following the standard practice of prior work \cite{mre19,scibert}, we used existing popular pre-trained BERT models as off-the-shelf tools and further fined-tuned these models with each of our corpora to build classifiers for scientific relation classification. Our investigation primarily concentrated on the benefits of fine-tuned BERT-based models brought to the classifiers. There exist some open questions regarding the characteristics of BERT's or SciBERT's pre-training settings and their influences on the downstream application. Example questions could be: (1) how is the quality of the corpus used to pre-train BERT; and, (2) how does the influence of this factor on BERT-based classifiers for identifying scientific relations? These issues are worthwhile to be explored in future. In addition, in this study, we mainly focused on the corpora covering AI-related research areas. The assessment of classifiers for identifying scientific relations in other domains need to be further studied.

\subsection{Potential Future Work}
To further assist digital library designers and librarians who want to build structural semantic representations over scholarly articles using scientific relation classifiers, there are three main avenues that are worthy of future exploration.

\subsubsection{Rethinking Human Annotation Guidelines}
Challenges in guaranteeing the quality of human annotations show that the process of defining scientific concepts and their relationships is still an open question, which can be further explored by experts. Researchers who are interested in this field can conduct their following studies in two directions. 

On the one hand, a formal understanding of the semantic relationships among terms in a specific research domain is critical to improve the clarity of annotation guidelines on this domain's scholarly information extraction. In particular, we suggest the consideration of semantic relation types should not only contain the semantic associations between scholarly terms, but also require the hierarchical structure of relation types. 

On the other hand, it would be helpful if the strategy for designing annotation guidelines could follow the ultimate application goal. In practice, building a scholarly knowledge graph can be used for  organizing general scholarly knowledge within a single or across multiple research domain(s) or managing knowledge with a specific focus, such as the evolution of scientific works' contributions. With different applications of scholarly knowledge graphs, the corresponding annotation rules on scientific terms and pre-defined relation types might be different. For example, annotations on the general scientific concepts shown in Figure~\ref{fig:kg} such as ``method", ``approach", ``system", and ``algorithm" are limited to informativeness for indicating the specific contributions of each research work.

\subsubsection{Benchmark Corpus from Multiple Domains}
While scholarly records in digital libraries usually cover various domains, publicly accessible corpora used for developing scholarly information extraction techniques primarily in several specific domains such as biology \cite{bio1,sie_survey} and computer science \cite{luan19,gabor2018semeval}. Given that scholarly publications in different domains may cover various relation types, and increasingly, language expressions such as word choice and writing style could be different, there is a demand for building a benchmark corpus consisting of scholarly records in a wide range of domains, which can be helpful in providing a comprehensive evaluation for the state-of-the-art relation classification techniques developed for extracting scholarly semantic information.

\subsubsection{Evaluation on Open Information Extraction Techniques}
In addition to building classifiers to identify pre-defined relations, techniques that are developed under the paradigm of open information extraction to identify more diverse relational triples ``without requiring any relation-specific human input'' ~\cite{openie} can an alternative yet promising strategy for extracting scholarly information, especially for identifying scientific relational tuples from scholarly records in various domains with massive uncertain semantic relations in advance. The further examination of state-of-the-art techniques in this field for scientific relation identification and the trade-offs between such techniques with \textsc{BERT}-based classification on pre-defined relations could be a valuable avenue to pursue.  

\section{Conclusions}
We have investigated the scientific relation classification task to support building scholarly knowledge graphs based on digital library collections. We provide a comprehensive view of eight \textsc{Bert}-based classification models on three clean corpora, which differ usefully in terms of corpus size and annotation guidelines. Moreover, considering many scholarly records in real-world digital libraries are digitized with OCR errors, we further prepared three noisy corpora corresponding to the clean ones and investigated the effect of OCR errors on the optimal \textsc{Bert}-based classifier identified from clean data. The presented empirical study in this paper contributes to the digital library stakeholder's understanding of state-of-the-art NLP techniques for identifying semantic relations from scholarly publications, which can provide practical benefits for identifying the optimal NLP tool to build scholarly knowledge graphs of digital library collections.

Our observations indicate that the performance of classifiers on clean texts is mainly associated with two aspects. First, from the perspective of training algorithms, three main factors, including classification strategies, the pre-training corpus domain and vocabulary case, determine the optimal model to apply in practice. Second, with respect to the annotation of scientific relations for training, there are two key factors that influence the ability of a \textsc{Bert}-based classification model to identify each relation type: (1) the number of annotations of each relation type; and, (2) the regularity of each relation's syntactic context. With further exploration on OCR noise impacts, we found that text noise caused by digitization has an obvious negative influence on the performance of \textsc{Bert}-based classifiers when identifying scientific relations, especially when the ratio of noise is high (e.g., ~49\%). Comparatively, relations with more concrete semantic meaning such as \textsc{Usage} and \textsc{Result} are more beneficial with the classifiers' robustness to OCR noise than other relations, while relations showing concept hierarchy like \textsc{Compare} and \textsc{Hyponym-Of} are more likely to stimulate the vulnerability of classifiers to the noise.

The overall insights in this study suggest that the uncased \textsc{SciBERT}-based classification model built under SRC strategy is the optimal choice for scientific relation classification in general. Regarding the corpus with OCR noise, we suggest DL stakeholders employ noisy data to build classifiers because the heterogeneous nature of OCR noise in training data is helpful with the generalization ability of classification models for processing unseen data.

\begin{acknowledgements}
The authors would like to thank anonymous reviewers for their constructive comments on this paper. We also appreciate Alaine Martaus, Jacob Jett and Yuerong Hu from University of Illinois at Urbana-Champaign for their helpful paper proofreading.
\end{acknowledgements}

%
%

\bibliographystyle{splncs04}       
\bibliography{ijdl_submission.bib}   


\end{document}